\documentclass[twocolumn, deluxetables]{aastex631}
\usepackage{amsmath}
\usepackage[caption=false]{subfig}
\usepackage{multirow}
\usepackage{graphicx}
\usepackage{xcolor}
\usepackage[T1]{fontenc} 
\usepackage{enumitem}
\usepackage{comment}
\usepackage{CJK}

\shorttitle{XRISM Spectroscopy of Winds in NGC 4151}
\shortauthors{Xiang et al.}

\graphicspath{{./}{figures/}}

\begin{document}

\title{XRISM Spectroscopy of Accretion-Driven Wind Feedback in NGC 4151}

\author[0000-0002-7129-4654]{Xin Xiang}
\affiliation{Department of Astronomy, University of Michigan, MI 48109, USA}
\email{xinxiang@umich.edu}

\author[0000-0003-2869-7682]{Jon M. Miller}
\affiliation{Department of Astronomy, University of Michigan, MI 48109, USA}

\author[0000-0001-9735-4873]{Ehud Behar}
\affiliation{Department of Physics, Technion, Technion City, Haifa 3200003, Israel}

\author[0000-0003-2704-599X]{Rozenn Boissay-Malaquin}
\affiliation{Center for Space Sciences and Technology, University of Maryland, Baltimore County (UMBC), Baltimore, MD, 21250 USA}
\affiliation{NASA / Goddard Space Flight Center, Greenbelt, MD 20771, USA}
\affiliation{Center for Research and Exploration in Space Science and Technology, NASA / GSFC (CRESST II), Greenbelt, MD 20771, USA}

\author[0000-0003-2663-1954]{Laura Brenneman}
\affiliation{Center for Astrophysics | Harvard-Smithsonian, MA 02138, USA}

\author[0009-0007-5987-0405]{Margaret Buhariwalla}
\affiliation{Department of Astronomy and Physics, Saint Mary's University, Nova Scotia B3H 3C3, Canada}

\author[0000-0002-3687-6552]{Doyee Byun}
\affiliation{Department of Astronomy, University of Michigan, MI 48109, USA}

\author[0000-0002-1065-7239]{Chris Done}
\affiliation{Centre for Extragalactic Astronomy, Department of Physics, University of Durham, South Road, Durham DH1 3LE, UK}

\author[0009-0006-4968-7108]{Luigi Gallo}
\affiliation{Department of Astronomy and Physics, Saint Mary's University, Nova Scotia B3H 3C3, Canada}

\author{Dimitra Gerolymatou}
\affiliation{Department of Astronomy, University of Geneva, Versoix CH-1290, Switzerland}

\author[0000-0002-5075-7920]{Scott Hagen}
\affiliation{Centre for Extragalactic Astronomy, Department of Physics, University of Durham, South Road, Durham DH1 3LE, UK} 

\author[0000-0001-5540-2822]{Jelle Kaastra}
\affiliation{SRON Netherlands Institute for Space Research, Leiden, The Netherlands} 
\affiliation{Leiden Observatory, University of Leiden, P.O. Box 9513, NL-2300 RA, Leiden, The Netherlands}

\author[0000-0002-8108-9179]{Stephane Paltani}
\affiliation{Department of Astronomy, University of Geneva, Versoix CH-1290, Switzerland}

\author[0000-0002-6374-1119]{Frederick S. Porter}
\affiliation{NASA / Goddard Space Flight Center, Greenbelt, MD 20771, USA}

\author[0000-0002-7962-5446]{Richard Mushotzky}
\affiliation{Department of Astronomy, University of Maryland, College Park, MD 20742, USA}

\author[0000-0001-6020-517X]{Hirofumi Noda}
\affiliation{Astronomical Institute, Tohoku University, Miyagi 980-8578, Japan}

\author[0000-0002-4992-4664]{Missagh Mehdipour}
\affiliation{Space Telescope Science Institute, 3700 San Martin Drive, Baltimore, MD 21218, USA}

\author[0000-0002-2933-048X]{Takeo Minezaki}
\affiliation{Department of Physics, University of Tokyo, Tokyo 113-0033, Japan}

\author[0000-0002-5097-1257]{Makoto Tashiro}
\affiliation{Department of Physics, Saitama University, Saitama 338-8570, Japan}
\affiliation{Institute of Space and Astronautical Science (ISAS), Japan Aerospace Exploration Agency (JAXA), Kanagawa 252-5210, Japan}

\author[0000-0002-0572-9613]{Abderahmen Zoghbi}
\affiliation{Department of Astronomy, The University of Maryland, College Park, MD 20742, USA}
\affiliation{HEASARC, Code 6601, NASA/GSFC, Greenbelt, MD 20771, USA}
\affiliation{CRESST II, NASA Goddard Space Flight Center, Greenbelt, MD 20771, USA}

\begin{abstract}
The hottest, most ionized, and fastest winds driven by accretion onto massive black holes have the potential to reshape their host galaxies. Calorimeter-resolution X-ray spectroscopy is the ideal tool to understand this feedback mode, as it enables accurate estimates of physical characteristics needed to determine the wind's kinetic power.  We report on a photoionization analysis of five observations of the Seyfert-1.5 galaxy NGC 4151, obtained with XRISM/Resolve in 2023 and 2024.  In the Fe K band, individual spectra require as many as six wind absorption components.  Slow ``warm absorbers’’ (WAs, $v_{\mathrm{out}} \sim 100 - 1000~\mathrm{km~s^{-1}}$), very fast outflows (VFOs, $v_{\mathrm{out}} \sim 10^3~{\rm km}~{\rm s}^{-1} - 10^4~{\rm km}~{\rm s}^{-1}$), and ultra-fast outflows (UFOs, $v_{\mathrm{out}} \sim 10^4~{\rm km}~{\rm s}^{-1} - 10^5~{\rm km}~{\rm s}^{-1}$ or $0.033 - 0.33~c$) are detected simultaneously, and indicate a stratified, multiphase wind.  Fast and variable emission components suggest that the wind is axially asymmetric. All of the wind components have mass flow rates comparable to or in excess of the mass accretion rate, though the slowest zones may be ``failed’’ winds that do not escape.  Two UFO components have kinetic luminosities that exceed the theoretical threshold of $L_{kin} \geq 0.5\% L_{Edd}$ necessary to strip the host bulge of gas and halt star formation, even after corrections for plausible filling factors.  The bulk properties of the observed winds are consistent with magnetocentrifugal driving, where the density depends on radius as $n \propto r^{-1.5}$, but radiative driving and other mechanisms may also be important. Numerous complexities and variability require further analysis.
\end{abstract}


\section{Introduction} \label{sec:intro}
Active Galactic Nuclei (AGNs), signifying the active gaseous growth phase of supermassive black holes (SMBHs) in the centers of nearly every galaxy, are fundamental to understanding galactic evolution and the interplay between the central black hole and its environment. The strong correlations between SMBH mass and the properties of the host galaxies are well established through observational black hole scaling relations, including the black hole mass - stellar velocity dispersion relation ($M_\mathrm{BH} - \sigma$; \citealt{Kormendy_Ho_2013}) and the black hole mass - stellar mass relation ($M_\mathrm{BH} - M_{*}$; \citealt{Greene_2020}). 

SMBHs grow primarily through the accretion of gas \citep{Soltan_1982}, a process that releases significant amounts of energy \citep{Lynden-Bell_1969}, far exceeding the binding energy of the host galaxy's bulge. However, as black holes grow, they also communicate some of their accretion energy to their host environment, regulating their own growth and influencing the surrounding gas. Powerful, ionized winds \citep{King_Pounds_2015} are one of the most effective feedback mechanisms, especially if their kinetic power ($\dot{K}_E$) exceeds 0.5-5\% of the AGN Eddington luminosity. A coupling efficiency of 0.5\% may be sufficient to substantially suppress star formation under idealized conditions \citep{Hopkins_2010}, while 5\% represents the canonical benchmark for strong, galaxy-scale impact in numerical simulations \citep{Di_2005}. Understanding connections between accretion and winds, including launching radii, driving mechanisms, and total feedback, is essential to understanding black hole growth and evolution. Since much of the gas involved is highly ionized and originates close to the central engine, X-ray observations are vital for probing these phenomena.

Atomic absorption features in the UV and X-ray spectra of AGNs suggest the presence of mass outflows in the form of ionized gas. These observable outflows span a wide range in properties, from slow warm absorbers (WAs, first detected via blue-shifted O VII and O VIII absorption edges by \citealt{Halpern_1984}) to ultra-fast outflows (UFOs, first detected via highly blueshifted Fe K absorption lines by \citealt{Chartas_2002}). For a recent review of these outflows, see, e.g., \cite{Gallo_2023, Laha_Reynolds_2021}. 

Photoionization modeling of the ionized atomic lines and edges enables self-consistent measurements of outflow properties, including the projected radial outflow velocities $v_{\mathrm{z}}~(\mathrm{km~s^{-1}})$, the ionization parameter $\xi~(\mathrm{erg~cm~s^{-1}})$, and the column density $N_\mathrm{H}~(\mathrm{cm^{-2}})$. The ionization parameter is defined as $\xi = L/n r^2$ \citep{Tarter_1969} where $L$ is the ionizing luminosity of the source in the energy band 13.6 eV - 13.6 keV, $n$ is the electron number density of the gas, and $r$ is the distance from the BH. The distance at which winds are launched can diagnose how they are driven \citep{Gianolli_2024}. Different ionization states and velocities of the gas are likely to sample the different regions of wind components, since ionization depends on distance from the central engine. A positive correlation between ionization parameter and column density is observed in the X-ray regime \citep{Crenshaw_Kraemer_2012, King_Miller_2013}, signaling that X-rays sample the bulk of the gas in AGN outflows.

WAs are typically observed as blue-shifted absorption lines and edges from H-like and He-like ions of C, N, O, Ne, etc, and iron transitions from the L-shell to M-shell (e.g., the iron unidentified transition array, or UTA) in the soft X-ray spectra of nearly 65\% of Seyfert galaxies, with outflow velocity of $v_{\mathrm{out}} \sim 100 - 1000~\mathrm{km~s^{-1}}$ \citep{Blustin_2005, Laha_2016, McKernan_Yaqoob_Reynolds_2007, Reynolds_1997}. These soft X-ray WAs typically have low ionization parameters in the range of $\log{\xi} \sim 1-3$ \citep{Sako_2001, Laha_2016}. Multiple WA components have been detected in the same source, differing in their outflow velocities, ionization, and column densities \citep{Laha_2014, Detmers_2011}. Warm absorbers with higher ionization parameters can also be detected in the Fe-K shell band \citep{Reeves_2004, Young_2005}. The wide range of warm absorber properties suggests different launching radii and may require different driving mechanisms. Hydrodynamical simulations have modeled WAs as thermally and radiatively driven winds originating from the dusty torus, where X-ray heating and radiation pressure shape the dynamics and launch a structured, multi-phase flow; magnetic fields may also contribute in some scenarios \citep[e.g.,][]{Dorodnitsyn_Kallman_Proga_2008, Dorodnitsyn_Kallman_Proga_2008_II, Kallman_Dorodnitsyn_2019}. Many studies place WAs between the broad line region (BLR; $\sim$ pc) for high ionization components to the distant molecular torus ($\sim$ kpc) \citep{Blustin_2005, Laha_2016, Krolik_Kriss_2001}. The kinetic power in an outflow is a steep function of velocity ($\dot{K}_E \propto v^3$), so WAs may not deliver strong feedback. However, they may help to transport angular momentum and may partly enable disk accretion.

UFOs are highly ionized winds observed as relativistically blue-shifted ($v_{\mathrm{out}} \sim 0.03 - 0.3~c$) Fe K-shell absorption lines from Fe XXV He$\alpha$ (6.70 keV) and Fe XXVI Ly$\alpha$ (6.97 keV), nominally in more than 40\% of Type-1 Seyferts and quasars based on low-resolution CCD spectra. They are observed in both near-Eddington sources (e.g., \citealt{Nardini_2015, Zoghbi_2015, Reeves_2018}) and sub-Eddington sources (e.g., \citealt{Zak_2024, Parker_2017}).  Like WAs, UFOs are observed with a range of properties:  ionization parameters in the range of $\log{\xi} \sim 3-6~\mathrm{erg~cm~s^{-1}}$, column densities between $N_\mathrm{H} \sim 10^{22} - 10^{24} \mathrm{cm}^{-2}$, and derived launching radii between $r = 10^{2-4} GM/c^2$ \citep{Tombesi_2011, Gofford_2015, Igo_2020}.  Since $\dot{K}_E \propto v_{\mathrm{out}}^3$, UFOs can carry considerable kinetic power and significantly influence the evolution of the host bulge, even halting star formation and eventually SMBH growth \citep{Tombesi_2010, Gofford_2015}. UFOs are likely transient, and difficult to detect for many reasons, but they may be at the heart of key observed galactic scaling relationships. Understanding how, why, and when UFOs are driven is critical to fully understanding the influence of black hole accretion on galaxies.  

The mechanisms that drive the launching of these winds are a topic of ongoing debate, with potential contributors including thermal pressure, radiation pressure, magnetic fields from the disk, or a combination of these (see, e.g., \citealt{Proga_2007}): 

Thermal winds \citep{Begelman_1983} are produced when X-rays from the central engine raise the temperature of the the outer disk surface up to the Compton temperature, so that the thermal velocity of the gas exceeds the local escape velocity. This mechanism may act in stellar-mass black holes, wherein the disk temperature is very high (\citealt{Done_2017, Tomaru_2022}; however, see \citealt{trueba2019}).  Below the Eddington limit, thermal winds are likely restricted to large radii where the escape velocity is naturally low; recent studies suggest a correspondingly low velocity limit of $v \simeq 200~{\rm km}~{\rm s}^{-1}$ \citep{higginbottom2017}.  Thermal winds may explain low-velocity WAs in some Seyfert galaxies \citep{Chelouche_2005}; however, thermal driving alone cannot explain the extreme velocities measured in UFOs. 

Radiatively driven winds can be accelerated by radiation pressure from the continuum or pressure from spectral lines. A continuum-driven wind is likely important at or above the Eddington limit, when the radiation can easily couple to highly ionized gas via electron scattering and naturally drive outflows to very high speeds $\sim 10,000~\mathrm{km~s^{-1}}$ (e.g. in some luminous quasars: \citealt{Nardini_2015, Reeves_2018} and some Narrow Line Seyfert 1: \citealt{Hagino_2016, Parker_2017, Meena_2021}). 
In contrast, radiation pressure on lines is most effective when the gas is moderately ionized -- within an ionization window where gas opacity is higher than the electron scattering, resulting in a strong force multiplier, $M(\xi, \tau)$ \citep[first studied by][]{Castor_1975}. The force multiplier can remain larger than unity out to $\log{\xi} \sim 3$ \citep{Dannen_2019}; however, the efficacy of line force is low ($M(\xi, t)$ drops to zero) for ionization parameters of $\log{\xi}> 3$.  Outflows driven by radiation can originate close to the central engine if self-shielding avoids over ionization of the gas \citep{Proga_2000, Proga_Kallman_2004}. 

Magnetic fields are likely central to the operation of accretion disks, whether through mass and angular momentum transfer along poloidal field lines \citep{Blandford_Payne_1982}, or via internal viscosity via the magnetorotational instability (or, MRI; \citealt{Balbus_Hawley_1991}).   Magnetic winds can operate over a broad range of radii, including at small radii where radiation forces on lines and thermal driving may be ineffective or impossible.  
Therefore, magnetically driven winds may be a key means by which UFOs are launched (\citealt{Blandford_Payne_1982, Contopoulos_Lovelace_1994, Ferreira_Pelletier_1993, Everett_Ballantyne_2004, Fukumura_2010}). Particularly in sub-Eddington sources and scenarios where line driving is inefficient, magnetic driving may be the dominant wind launching mechanism.  
Several observations may favor magnetic driven winds in both AGNs \citep{Fukumura_2015, Fukumura_2018, Kraemer_2018} and XRBs \citep{Miller_2006, Fukumura_2017}.  Some simulations indicate that magnetic driving mechanisms and radiation force can also work jointly accelerate outflows \citep{Proga_2003, Everett_2005, Ohsuga_Mineshige_2011, Fukumura_2015, Nardini_2015}.

Different launching mechanisms are likely to result in different absorption line profiles \citep{Gallo_2023}, as well as different density profiles.  Early treatments of magnetic winds predict $n(r) \propto r^{-1.5}$, for instance, while later models with different assumptions predict flatter profiles, consistent with $n(r) \propto r^{-1}$ (\citealt{Blandford_Payne_1982}, \citealt{Contopoulos_Lovelace_1994}).  The Absorption Measurement Distribution (AMD; \citealt{Holczer_2007, Behar_2009}) makes the density distribution accessible through direct observables.  The AMD is defined as the distribution of the column density as a continuous function of $\log{\xi}$, $\mathrm{AMD} \equiv dN_{\mathrm{H}} / d \log{\xi}$. The AMD slope -- obtained by parameterizing the AMD as a power law -- has a direct connection with the density profile $n(r)$ of the outflow. 

Although low-ionization WAs are well-studied at low-energy in deep observations with Chandra and XMM-Newton (e.g, \citealt{Gallo_2023} ) and although UFOs are seen in a subset of high-Eddington AGN (e.g., \citealt{Reeves_2016, Lobban_2016}), there are few examples of both slow and fast winds in the Fe~K band (see, e.g., \citealt{Laha_2016}).  The presence or absence of ``very fast outflows'' (or, VFOs) with velocities between WAs and UFOs, is also uncertain.  Understanding the physical demographics of highly ionized outflows is an important step toward a full understanding of black hole feedback.  The resolution and sensitivity of XRISM \citep{Tashiro_2025, Ishisaki_2022} enable this crucial next step.  XRISM observations of PDS~456 strongly confirm UFOs in this extreme source \citep{XRISMPDS456_2025}, and reveal rich Fe~K band absorption in NGC 4151 at $\lambda_{Edd} \simeq 0.04$ \citep{XRISM_NGC4151_2024}.  

NGC 4151 is a particularly bright and proximal Seyfert 1.5 galaxy ($z = 0.0033, M_{\mathrm{BH}} = 3.4^{+0.4}_{-0.4} \times 10^7 M_{\odot}$; \citealt{Bentz_2015}). It has been extensively studied across multiple wavebands for decades \citep{Giacconi_1974, Holt_1980, Code_1982, Evans_1993, Maisack_1993, George_1998}.  NGC 4151 shows strong multi-wavelength variability from optical, to UV, and X-rays \citep{Ives_1976, Ulrich_1997, Edelson_2017}.  In most observations, the bolometric flux implies an Eddington fraction of $\lambda_{Edd} = 0.01-0.02$ \citep{Couto_2016}.  The optical narrow line region (NLR) appears to be inclined at an angle of $\theta = 45^{\circ}$ with respect to our line  of sight \citep{Das_2005}; this may indicate that our line of sight just skims the surface of the distant molecular torus, potentially enhancing variability through changing obscuration.
Above 2~keV, the continuum spectrum of NGC 4151 can be described by a power-law with a photon index of  $1.2 \leq \Gamma \leq 1.9$ \citep{Ives_1976, Beckmann_2005}; in some phases, the continuum is flatter than typical values observed in Seyfert-1 AGN ($\Gamma \sim 1.7$), likely signaling contributions from reflection. The power-law has a cut-off at around 100~keV \citep{Jourdain_1992, Fabian_2015}. The narrow Fe K$\alpha$ fluorescence line at 6.4 keV in NGC 4151 is one of the brightest observed in this class; recent studies with Chandra and XRISM show that it reveals the disk, broad line region, and torus \citep{Miller_2018, XRISM_NGC4151_2024}. 

The full X-ray spectrum of NGC 4151 reveals complex absorption due to multiple layers of neutral absorbers and ionized winds \citep{Kraemer_2020}.  Observations with Chandra clearly reveal He-like and H-like Fe XXV and Fe XXVI absorption lines in the Fe K band \citep{Couto_2016}.  These absorption and emission components likely span a broad range in radius, from the inner part of the BLR \citep{Bentz_2006}, through the intermediate line region (or ILR; \citealt{Kraemer_2006}), out to the outer dusty torus in the NLR \citep{Radomski_2003}. In addition, potential UFOs have been detected in some observations of NGC 4151 at CCD resolution \citep{Tombesi_2010}.  These unique characteristics -- variability, multi-layer absorption, and prominent outflows -- make NGC 4151 an exceptional laboratory for the study of wind feedback in AGNs.  In this paper, we present an analysis of the rich absorption spectra revealed in five observations of NGC 4151 with XRISM/Resolve. Section \S\ref{sec:data} describes the data reduction process for the XRISM observations. In Section \S\ref{sec:results}, we outline the spectral fitting methods and present the fitting results. The properties of the winds are calculated and discussed in Section \S\ref{sec: wind properties}. We further explores the implications of these findings in Section \S\ref{sec:discussion}, followed by Section \S\ref{sec: summary}, which summarizes our conclusions.

\section{Observation and Data Reduction} \label{sec:data}

\begin{deluxetable*}{cccccc}
\tablecaption{Observation log for the XRISM/Resolve of NGC 4151}
\label{table:observationlog}
\tablewidth{0pt}
\tablehead{
\colhead{Obs} & \colhead{obsID} & \colhead{Start Time} &  \colhead{MJD} & \colhead{Exposure} & \colhead{$L_{\mathrm{ion}}$} \\ 
& & (UT) & & (ks) & ($10^{43}~\mathrm{erg~s^{-1}}$)
}

\startdata
    1 & 000125000 & 2023-12-02 06:22:50 & 60280 & 72 & $2.85^{+0.04}_{-0.08}$\\
    2 & 000137000 & 2023-12-26 18:32:51 & 60305 & 56 & $4.00^{+0.02}_{-0.18}$ \\
    3 & 300047020 & 2024-05-18 08:25:21 & 60448 & 98 & $2.67^{+0.25}_{-0.11}$ \\
    4 & 300047030 & 2024-06-16 00:58:31 & 60477 & 83 & $3.33^{+0.11}_{-0.05}$ \\
    5 & 300047040 & 2024-06-22 15:09:05 & 60484 & 89 & $3.22^{+0.27}_{-0.01}$ \\ \hline
\enddata
\tablecomments{$L_{\mathrm{ion}}$ is the 13.6 eV - 13.6 keV ionizing luminosity including all the emission and continuum components calculated from the best-fit models summarized in Table \ref{table:parameters}. The errors on ionizing luminosity are calculated based on the fractional error in the normalization of the power-law.}
\end{deluxetable*}

\subsection{XRISM}
Table \ref{table:observationlog} summarizes the information for the five XRISM/Resolve observations and the ionization luminosity using the best-fit models summarized in Table \ref{table:parameters} (detailed in \S\ref{sec:results}). The errors on the ionizing luminosity are calculated based on the fractional error in the normalization of the power-law (see below). The data reduction process is the same as detailed in the first XRISM paper on NGC 4151 \citep{XRISM_NGC4151_2024}. Hereafter, the five observations are referred to as Obs. 1, Obs. 2, Obs. 3, Obs. 4, and Obs. 5.

\subsection{NuSTAR} \label{sec: Nustar}
We also utilize a set of four NuSTAR observations to help define the broadband continuum spectrum of NGC 4151.  These were obtained simultaneously with XRISM Obs. 2--4.   Each NuSTAR observation had an exposure time of 46~ks. The corresponding ObsIDs are 60902010002, 60902010004, 60902010006, 60902010008; starting on 2023 December 27 at 10:36:09 (UTC), 2023 May 19 at 09:26:09 (UTC), 2023 June 15 at 12:11:09 (UTC), and 2023 June 22 at 04:21:09.  The data from each observation were reduced using the tools in HEASOFT version 6.34 and the corresponding NuSTAR calibration database (CALDB) files.  The ``nupipeline'' tool was run to reprocess the data using the latest calibration files.   Circular regions with radii of 120'' were used to select source and background regions on the FPMA and FPMB detectors.  Source and background spectra and responses for each observation were then generated by running the tool ``nuproducts.''  In each case, the spectral and response files from the FPMA and FPMB were combined using the tools \texttt{addascaspec} and \texttt{ftaddrmf}.

\section{Analysis and Results} \label{sec:results}
In this section, we present the results of spectral modeling of the five epochs of XRISM/Resolve observational data in the 2.4--17.4~keV range.  This gives a total fitting band of 15 keV, and avoids small lingering response uncertainties at low energy (close to the gate valve transmission threshold).  The fits focus on photoionization modeling (details in subsection \S\ref{sec: photoionization modeling}), on top of the base model that includes the emission and continuum components (details are described in \cite{XRISM_NGC4151_2024} and the following subsection \S\ref{sec: continuum modeling}).  All fits in this paper were made using SPEX version 3.08.01 \citep{Kaastra_1996}, minimizing a Cash statistic ($C$-stats; \citealt{Cash_1979}).  All of the errors in this work are based on the value of a given parameter on the boundary of its 1$\sigma$ confidence interval. Table \ref{table:parameters} lists all the best-fit parameter values and errors for the continuum and photoionization models.

\subsection{The continuum and narrow emission lines} \label{sec: continuum modeling}
We adopted a continuum model consisting of a blackbody component ``\texttt{$bb$}'' to represent the UV accretion disk, and a cut-off power-law component ``\texttt{$pow$}''.  We used two ``\texttt{$etau$}'' components in SPEX to bend the power-law to zero flux at both low (``\texttt{$etau_{low}$}'') and high (``\texttt{$etau_{hi}$}'') energies with cutoffs of 0.0136 keV and 300 keV, to ensure a realistic spectral shape and ionizing luminosity. The total continuum was further modified by the ``\texttt{$hot$}'' model to fit neutral partial-covering absorption, including the contribution from within NGC 4151 and the Milky Way (the Milky Way contributes only $N_\mathrm{H} = 2.1 \times 10^{20}$; \citealt{Collaboration_2016}) with two free parameters: the total equivalent hydrogen column density and the covering factor.  Despite its name, the default parameters of this model fit neutral partial-covering absorption, and this is its function in our model.  In addition, the redshift model ``\texttt{$reds$}'' is applied with a fixed source redshift of z = 0.0033.  This has the effect of shifting all emission and absorption components to the frame of NGC 4151.

To estimate the important FUV contribution to the ionizing luminosity, we considered the set of Hubble/STIS FUV spectra presented in \cite{Kraemer_2006}.  The spectra were obtained between 1999 and 2002, and span a broad range in mean flux.  Simple fits to the XRISM/Resolve spectra obtained in December 2023 indicate that the source flux ranked among the highest recorded, likely exceeding $F = 1\times 10^{-10}~{\rm erg}~{\rm cm}^{-2}~{\rm s}^{-1}$ in the 2--10~keV band.  Based on the close tracking between UV and X-ray flux trends observed in Swift monitoring observations \citep{Edelson_2017}, we elected to consider the STIS spectrum with the highest flux in the \cite{Kraemer_2006} study.  The E140m spectrum on 2002 May 8 was gathered over 7.6~ks, with observed flux densities as high as $F = 7.5\times 10^{-13}~{\rm erg}~{\rm cm}^{-2}~{\rm s^{-1}}$\AA$^{-1}$ at $\lambda = 1350$~\AA. Using the same STIS spectrum analyzed in \cite{Kraemer_2006}, we created a file that could be analyzed using XSPEC and SPEX via the FTOOL \texttt{ftflx2xsp}.  Assuming an intrinsic reddening of $E(B-V) = 0.04$ \citep{Kriss_1995} and a line-of-sight reddening of $E(B-V) = 0.027$ \citep{Schlegel_1998}, we fit the STIS spectrum with a simple disk blackbody function \citep{Mitsuda_1984}.  Regions of the spectrum with strong emission and absorption lines were ignored.  A statistically acceptable fit is not possible, but the disk blackbody describes the flux trend adequately.  The best-fit disk has a temperature of kT $= 16.8$~eV ($T = 1.9\times 10^5~\mathrm{K}$) and  gives an ionizing luminosity of $L_{\mathrm{ion}} = 3.2\times 10^{43}~{\rm erg}~{\rm s}^{-1}$ (0.0136--13.6~keV).  For a quick comparison, simple fits to Obs. 2 imply an X-ray luminosity of $L = 7.8\times 10^{42}~{\rm erg}~{\rm s}^{-1}$ (2--10~keV).

\begin{figure}
\includegraphics[width=0.48\textwidth]{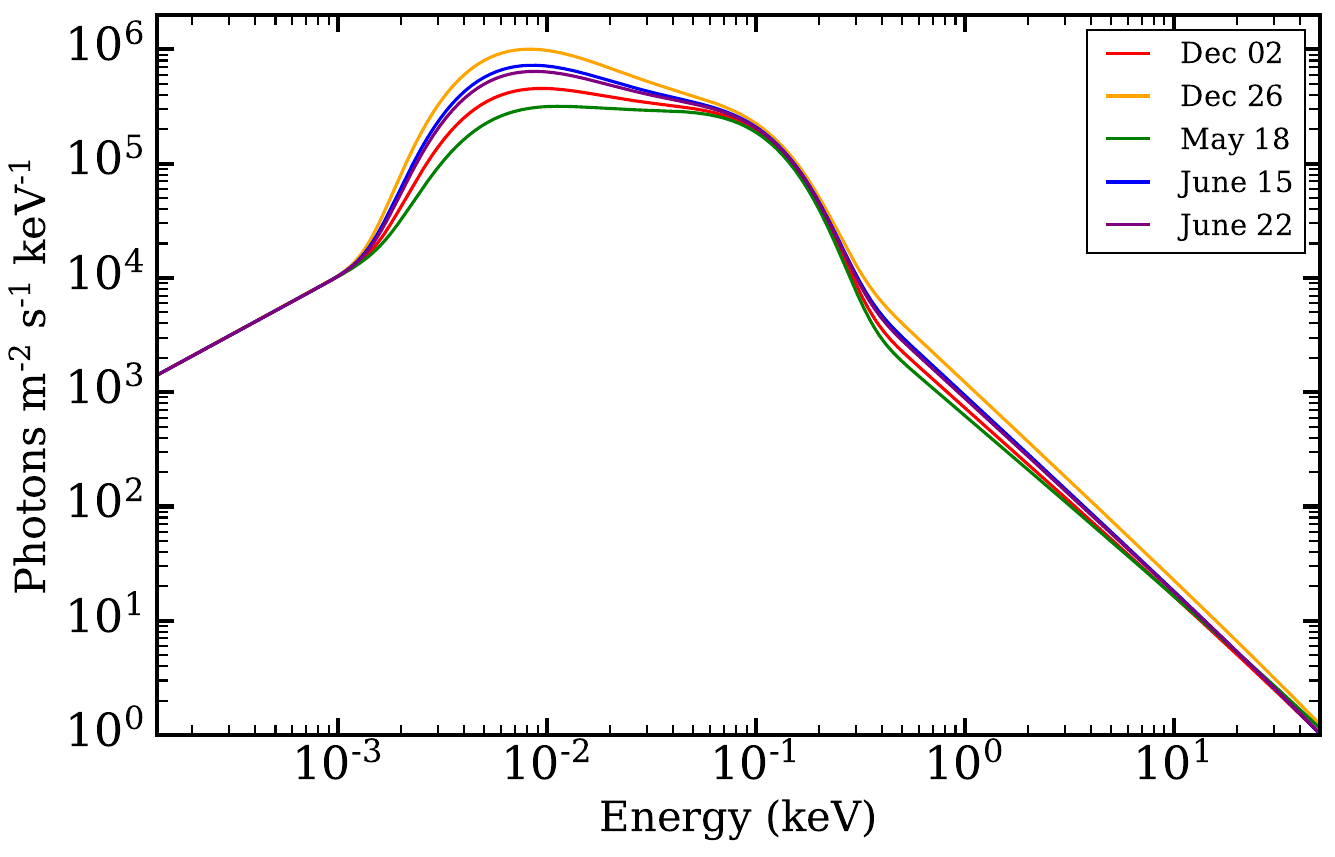}
\caption{Spectral energy distribution (SED) models for the five NGC 4151 observations, each including a low- and high-energy cutoff power-law and a blackbody component.}
\label{fig: SED}
\end{figure}

Clearly, it is important to include the FUV contibution to the ionizing SED when modeling the X-ray features observed with XRISM/Resolve.  Within SPEX, we defined a disk component by using ``\texttt{$bb$}'' with a fixed temperature of kT = $2 \times 16.8$ eV at the peak emissivity. We manually adjusted the emitting area and fixed it at $7.5 \times 10^{20}~\mathrm{m^2}$ so that the disk component creates a similar amount of ionization luminosity to the STIS spectrum and fed this emission into the total ionization luminosity in addition to the X-ray continuum components, which will be important to calculate the ionizing balance for the photoionization modeling in Section \S\ref{sec: photoionization modeling}.  We note that the FUV emission may not be a simple blackbody or disk blackbody function, but rather the result of Compton up-scattering from a cooler disk \citep{Edelson_2017}.  For the purposes of building a credible ionizing spectral energy distribution (SED), however, these approximations are expedient and unlikely to introduce significant errors. Figure \ref{fig: SED} shows the SED model for each observation containing the best-fit cutoff power-law and blackbody components.

We next performed an initial continuum fit over the full 2.4--17.4~keV band with a binning factor of 45 that combined every 45 channels, to constrain the following free parameters: the power-law index ($\Gamma$), normalization, and neutral absorption column density and covering factor. The parameters of the blackbody component describing the accretion disk were held fixed in this fit and all subsequent fits.  Ignoring the lines in this manner did not yield an acceptable fit, of course, but provided a basis for building a better model.

The neutral Fe K$\alpha$ emission line was modeled using three ``\texttt{$mytorus$}'' line components \citep{Murphy_Yaqoob_2009}.  Each of these components is a table model (\texttt{mytl\_V000010nEp000H100\_v00.fits}), converted into a format compatible with SPEX and broadened by an independent ``\texttt{$spei$}'' model. The details of Fe K$\alpha$ modeling are described in \cite{XRISM_NGC4151_2024}. Although the emission lines are not the primary focus of this paper, we included the Nickel K$\alpha_1$ and K$\alpha_2$ lines, with their widths linked, using the ``\texttt{$line$}'' model in SPEX.  This ensures that residuals in the band surrounding 7.4~keV are not influenced by unmodeled Ni~K$_{\alpha}$ line flux. The free parameters for the narrow-band emission line fit were as follows, (1) from ``\texttt{$mytorus$}'' model: the column density of the emitting gas ($N_H$), the inclination of the emitting gas ($i$), and the flux normalization; (2) from ``\texttt{$spei$}'': the inclination (linked to the ``\texttt{$mytorus$}'' model) and the inner radius ($r$); (3) from ``\texttt{$line$}'': the optical depth ($\tau$) and the FWHM of the Gaussian and Lorentzian component. The continuum and narrow emission lines model can be summarized as:
\begin{multline}
     (bb + (pow * etau_{hi} * etau_{low} + 3 \times (mytorus * spei))) \\ 
     * line_1 * line_2 * hot * reds
\end{multline}

The resolution and sensitivity of Resolve can make it difficult to simultaneously fit the continuum and detailed line structure within a narrow band, so we adopted an iterative approach.  We first focused on the broadband fit over the full 2.4--17.4~keV band to constrain the continuum parameters.  After obtaining a preliminary continuum model, we froze the continuum parameters and applied the emission line models in a more focused fit to the 5.4--10.4~keV band.  The data were binned by a factor of 15 in the 5.4--7.4~keV band, and 45 in the 7.4--10.4~keV band, based on their relative signal and the widths of evident features (native Resolve bins have a width of 0.5 eV).  After several iterations of these steps, all continuum and line parameters were allowed to vary simultaneously in a fit across the full 2.4--17.4~keV band.  (The upper panel of Figure \ref{fig: brd} presents the five XRISM spectra and corresponding best-fit models in the broad 2.4--17.4~keV range, including the absorption models described below.)  This process yielded the preliminary best-fit parameters for both the continuum and narrow emission lines, providing a foundation for incorporating photoionization models -- the primary focus of this paper -- in subsequent steps.

\subsection{Detailed Photoionization Modeling and Results} \label{sec: photoionization modeling}

The high-resolution XRISM/Resolve spectra of NGC 4151 reveal complex wind absorption (and emission) features from different ionization states of Fe.  We employed the ``\texttt{$pion$}'' photoionization model in SPEX \citep{Miller_2015, Mehdipour_2016} to self-consistently describe these features. \texttt{$Pion$} calculates the ionization balance using the ionizing radiation from the continuum components in SPEX, which can vary between the absorber layers. At each step of the fitting process, the parameters of the ionizing continuum are determined simultaneously with the parameters of the absorbing layers. Additionally, the outer absorber layers see the transmitted luminosity from the inner layers instead of seeing the full central continuum, providing a physically realistic treatment of layered absorption. 

The free parameters for the ``\texttt{$pion$}'' absorption components are: (1) the column density ($N_H~(\mathrm{cm^{-2}})$), (2) the ionization parameter ($\xi~(\mathrm{erg~cm~s^{-1}})$), (3) the bulk velocity of the gas in the source frame along our line-of-sight ($v_z~(\mathrm{km~s^{-1}}$), negative for blue-shifted components), and (4) the turbulent velocity of the gas ($\sigma_z~(\mathrm{km~s^{-1}})$). In all fits, the geometric covering factors for the absorbers are assumed to be 0.5, which is appropriate since approximately half of AGNs have ultra-fast, high-ionization outflows \cite{Tombesi_2010}. All other ``\texttt{$pion$}'' parameters retained their default values, including elemental abundances, which were kept at unity relative to solar values in all fits.  The multi-layers were structured sequentially, with the absorbed continuum from the inner layer feeding into the next ``\texttt{$pion$}'' component.  We do not fit for a parameter called ``lixi'' in each component ($L_{\mathrm{ion}}/\xi$, useful for tracing the flux that is incident at the inner edge of each layer); this parameter is automatically calculated by the model.  

These fits were also made over the narrow energy band of 5.4--10.4~keV, again with a binning factor of 15 between 5.4--7.4~keV and 45 between 7.4--10.4~keV.  The stronger binning at higher energy particularly enabled searches for broad UFOs in absorption. After a detailed initial assessment, we identified that each observation may require up to four layers of WAs and VFOs, and up to two layers of UFOs.  Not every observation requires six absorption zones at high statistical significance, but applying a {\em consistent} model framework to {\em every} spectrum is the only means of tracing how wind components evolve over time, in response to ionizing flux, etc.   Indeed, it is readily apparent that the number, strength, and statistical significance of the absorption layers varies from observation to observation.  In addition to these absorption zones, we detected transient re-emission from the wind -- sometimes blue-shifted -- in specific observations; these required dedicated ``\texttt{$pion$}'' emission components.  

Figure \ref{fig: diagram} provides a schematic summary of the full modeling framework that we adopted to fit the Resolve spectra of NGC 4151.  Figure \ref{fig: dec02}, \ref{fig: dec26}, \ref{fig: may18}, \ref{fig: june15}, and \ref{fig: june22} show the spectra and best-fit model in the energy band of $5.4-10.4$ keV with zoom-in panels in $6.45-7.05$ keV for Obs. 1, Obs. 2, Obs. 3, Obs. 4, and Obs. 5 respectively.

There are numerous ways to estimate the significance of different model components.  For single spectral lines, the flux normalization of the line divided by its $1\sigma$ minus-side error approximates the Gaussian-equivalent sigma at which the line is required.  In the case of wind components, which consist of many lines, their overall strengths are primarily governed by the column density.  Thus, dividing the best-fit column density by its $1\sigma$ minus-side error is a similar estimate of the significance at which a given component is required by the data.  Using such methods, the vast majority of the wind components in Table \ref{table:parameters} are highly significant.  However, this is not the most conservative procedure, as it does not fully account for potential degeneracies between components (e.g., multiple wind components contributing to a single line, and potentially diminishing it rather than augmenting it).  For similar reasons, even tracking changes in the Cash statistic or the Akaike Information Criterion (AIC; \citealt{Akaike_1974}) after adding components is not fully conservative, as the existing components may adjust in response to the addition of components, complicating direct comparisons. To address this, we introduce two complementary estimators of the significance of each wind component, though it remains informative to also examine column densities and their uncertainties.

In both cases, the method is accomplished by removing a given component from the total model, refitting, and utilizing the difference in the fit statistic.  First, we calculated a Gaussian equivalent sigma value based on the p-value derived from the change in Cash statistic and the number of degrees of freedom.  This method is common within the field, but it may fail to accurately capture significance when improvements are modest and/or when models are complex.  Therefore, we also calculated the change in the AIC when single components are removed, where $\rm AIC = n \ln{(C / n)} + 2p$ ($C$ is the Cash statistic value of the fit, $n$ is the number of data bins, and $p$ is the number of free parameters in the model.   A model with a lower AIC value represents an improvement to the fit.  As a guideline: a difference within 2 AIC units suggests both models fit the data at least equally well, a difference of 2 or more AIC units indicates substantial support for the model with the lower AIC, a difference of 10 or more AIC units suggests that the model with the higher AIC fails to explain substantial features of the data \citep{Burnham_2002}. We listed the $\Delta AIC$ that is calculated by subtracting the AIC value of the best-fit model from the model without a specific component. This implies that if $\Delta \rm AIC < -10$, the component must be required to explain the absorption/emission features; if $\Delta \rm AIC < -2$, the component has substantial support; if $-2 < \Delta \rm AIC < 2$, the component is marginally required; if $ 2 < \Delta \rm AIC < 10$, the model without the component is substantially supported; if $\Delta \rm AIC > 10$, the additional of the component is likely false. The best-fit parameters, detection significance (D.S), and $\Delta \rm AIC$ for each component are listed in Table \ref{table:parameters}.  A detailed description of the fitting procedures and results follows below.

\subsubsection{WA and VFO components: wind zones \#1--4}
In every observation, a sequence of narrow absorption lines is resolved in the 6.45--6.65~keV band, and readily identified as Fe XX--XXIV (at 6.50 keV, 6.54 keV, 6.59 keV, 6.63 keV, 6.66 keV, 6.70 keV, respectively) with small ($\sim 100~{\rm km}~{\rm s}^{-1}$) blue-shifts.  More prominent Fe XXV (He$\alpha$ at 6.70 keV, He$\beta$ at 7.88 keV) and Fe XXVI (Ly$\alpha_1$ and $\alpha_2$ at 6.95 and 6.97 keV, Ly$\beta$ at 8.25 keV) absorption lines are observed simultaneously.  No fit with a single ionization zone could reproduce the range of observed charge states, nor the observed line ratios.  In the most complex spectra, three layers of slow WAs were necessary to fully describe the complex absorption lines from Fe XX--XXVI within the wind.

In all WA layers, the turbulent velocity was allowed to vary in the range  $100~{\rm km}~{\rm s}^{-1} \leq \sigma_{z} \leq 1000~{\rm km}~{\rm s}^{-1}$.  Significant re-emission is most likely from wind components with high column densities.  The WA component that produced the strongest Fe XXV and Fe XXVI absorption lines was therefore coupled with a re-emission component, linking the values of the column density, ionization parameter, and turbulence velocity.  Due to the uncertain geometry of the re-emitting regions, the covering factors of the emitters are left free, and the emission was further broadened using a Gaussian function (via convolution with the ``\texttt{$vgau$}'' model in SPEX).  Table \ref{table:parameters} details the full best-fit model for each observation.  Within this table, the warm absorbers are marked as \#1--3 step progressively closer to the continuum components (the central engine).  The re-emission coupled with ``\texttt{$pion_{\#2}$}'' was labeled ``\texttt{$pion_{emis\#2}$}''.  For Obs. 2, the fitting regressed to zero column density for this re-emitter, so this component was omitted for that observation.  In all spectra, absorption is dominant over re-emission, indicating relatively little gas above our line of sight.

The wind component that is layered farthest from the central engine, \#1 (``$pion_{\#1}$''), tends to have lowest ionization parameter -- around $\log\xi = 2.6-2.7$ -- producing a high ratio of absorption features from lower ionization states of Fe. These WA components are marked as green curves in Figure \ref{fig: dec02}, \ref{fig: dec26}, \ref{fig: june15}, and \ref{fig: june22} with velocity in a range of $\sim 100 - 400~\rm km~s^{-1}$. However, in Obs. 3, ``$pion_{\#1}$'' has a best-fit velocity of $v_z = -3092^{+46}_{-47}~\rm km~s^{-1}$ with D.S = $3~\sigma$, and is therefore classified as a VFO. It has a low ionization parameter and captures the blue-shifted Fe XXI - Fe XXV absorption, shown as the green curve in the zoom-in panel of figure \ref{fig: may18}.   In the other four observations, the properties of a given wind component are fairly consistent.  The significant change in outflow velocity between Obs. 2 and Obs. 3 -- despite similar ionization parameters and ionizing luminosities -- may indicate a short-term acceleration of the wind that cannot be explained via radiation pressure, or/and new gas crossing our line of sight. 

Component\#2 (``$pion_{\#2}$'') is layered inside of component\#1, and it produces strong Fe XXV and Fe XXVI absorption features, marked as the orchid curves in Figure \ref{fig: dec02}, \ref{fig: dec26}, \ref{fig: june15}, and \ref{fig: june22}.  The best fit velocities across five observations are $v_z = -298^{+28}_{-29}, -288^{+32}_{-30}, -311^{+24}_{-23}, -266^{+22}_{-21}, -264^{+23}_{-23}~\rm km~s^{-1}$.  Although the velocities are fairly consistent, varying by less than 6\%, the column density and ionization parameter of this layer varies by $33\%$ and $0.37$ dex, in the same direction.  In addition, a $19\%$ variation in turbulence velocity is observed.  This strongly suggests that although component\#2 likely traces the same outflow zone across all five observations, the internal conditions of the gas are dynamic.  Such changes may reflect small variations in the filling factor, small variations in ionization, or/and transverse motion across line-of-sight.  Components\#2 exceeds the $> 3\sigma$ threshold in each observation and yields large changes in the AIC.  The re-emission features ($pion_{emis\#2}$) associated with the WA component\#2 (``$pion_{\#2}$'') have outflow velocities ranging from $\sim 400 - 1000~km~s^{-1}$.  Note that as these components are seen in emission, the outflow velocity is a red-shift. These components are not strongly required by the p-value estimator but are favored by the AIC.  The fact that the overall wind spectrum in each observation is strongly dominated by absorption indicates that the wind does not have a significant geometrical extent above our line of sight.

The component\#3 (``$pion_{\#3}$'') has diverse properties across the five Resolve observations (marked as blue curves in Figure \ref{fig: dec02}, \ref{fig: dec26}, \ref{fig: may18}, \ref{fig: june15}, and \ref{fig: june22}).   The columns, ionization parameters, and particularly the outflow velocities vary considerably.  This component ranks as the fastest WA in Obs. 1 and 2, but as the slowest WA in Obs. 3.  In Obs. 4 and Obs. 5, this component qualifies as a VFO, though it is worth noting that the speeds in these observations only slightly exceed that in Obs. 2.  In Obs. 3, this component primarily contributes lines from Fe XXI-XXIV.  However, for Obs. 4 and Obs. 5, this component primarily contributes to absorption in Fe XXV and Fe XXVI.  They are similar to component\#2 in terms of ionization, but they have significantly higher velocities.  If component\#3 changes into component\#2 as a continuous wind reaches a larger radius, it must decelerate significantly in every case except Obs. 3.  
Here again, the WA structure of Obs. 3 is different.  Swapping the component\#1 and component\#3 in Obs. 3 does not significantly change the Cash statistic.  However, it would not make either component more consistent with likely counterparts in the other observations.  As with component\#1, then, component\#3 illustrates that components within WA vary significantly on timescales of weeks and months.   

\begin{figure*}
\includegraphics[width=\textwidth]{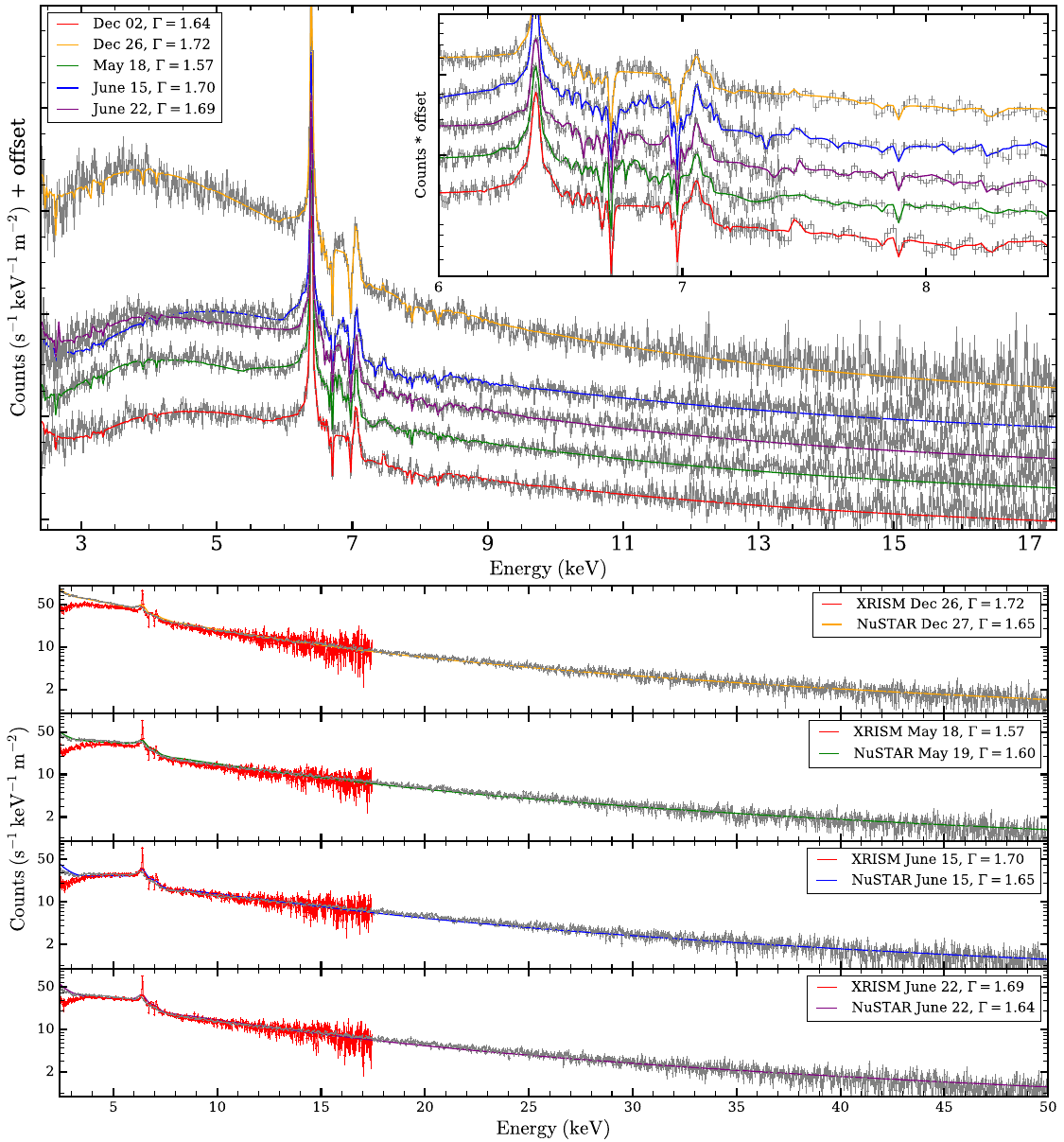}
\caption{UPPER PANEL: Five XRISM observations of NGC 4151 on Dec 02, Dec 26, May 18, June 15, and June 22, along with their respective best-fit models (colored curves) as detailed in Table \ref{table:parameters}, fitted separately in the 2.4–17.4 keV range. For better visualization, the flux of each observation is offset. The data are binned with a factor of 45 in the main panel and a factor of 15 in the zoomed panel (upper right) below 7.4 keV. LOWER PANEL: Four simultaneous NuSTAR observations in the 2-50 keV range refitted with the best model for XRISM observations with the power-law index, normalization free (colored curves). The best-fit power-law indices are within $\Delta \Gamma \leq 0.1$ and exhibit no significant deviations from the XRISM spectra.}
\label{fig: brd}
\end{figure*}

\begin{figure*}
\includegraphics[width=\textwidth]{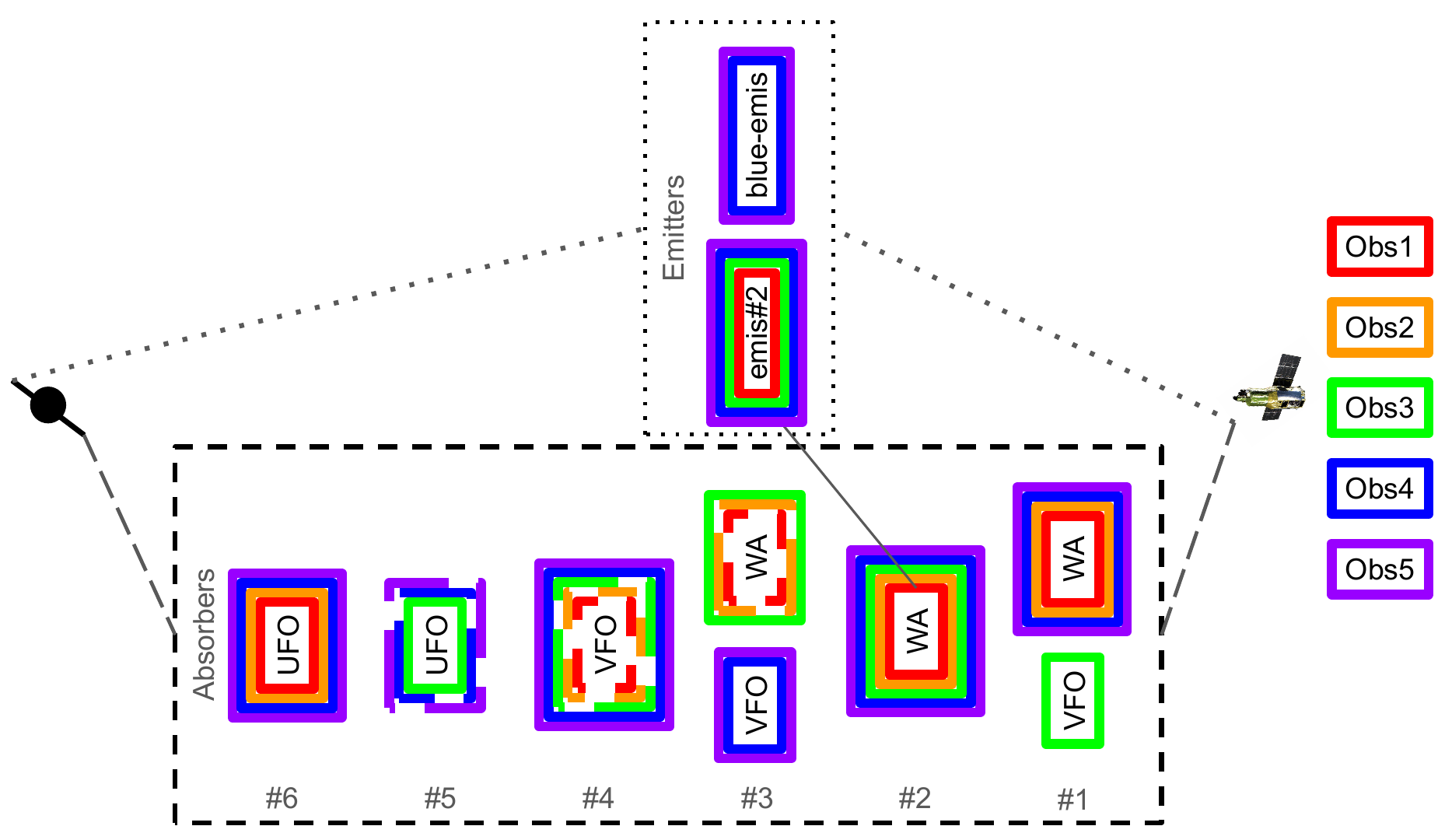}
\caption{Schematic representation of the multi-layer outflow model for five XRISM/Resolve observations of NGC 4151. The model consists of absorbers and emitters that create complex combinations of absorption and emission features. The absorbers are composed of up to four layers of WAs or VFOs, along with two UFO layers. Color-coded solid and dashed boxes indicate the presence of specific outflow components with strong ($\sigma >=3$) and weak ($\sigma < 3$) detection significance in different observations. Emitters include re-emission from the WA component \#2 and a blue-shifted emission feature.}
\label{fig: diagram}
\end{figure*}

\begin{figure*}
\includegraphics[width=\textwidth]{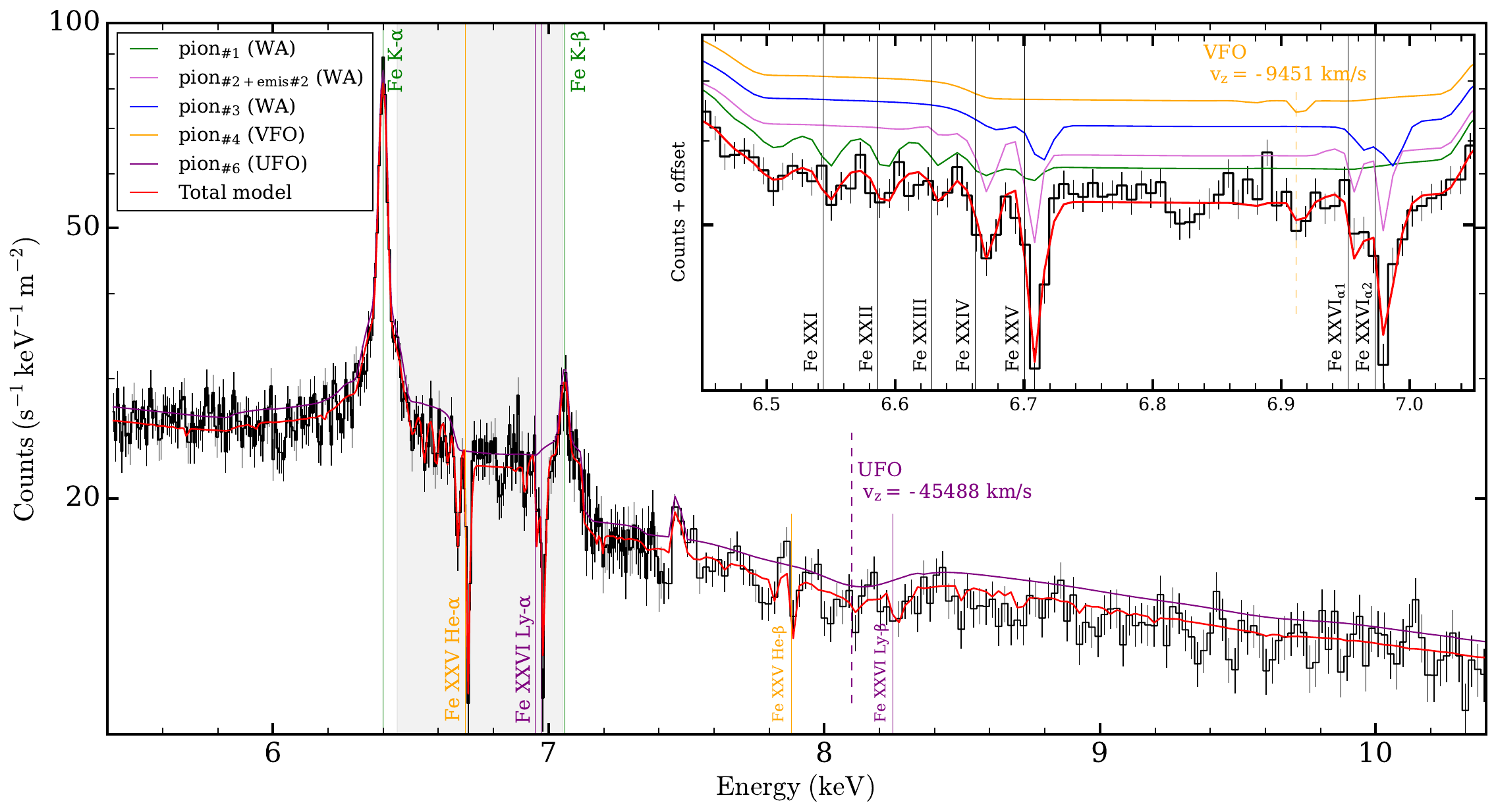}
\caption{XRISM observation of NGC 4151 on Dec 02 (Obs. 1). The best-fit model is shown as the red curve in both the main panel and zoom-in panel (upper right). Solid vertical lines marked the absorption/emission features for corresponding Fe states at lab values. In the zoom-in panel, the components for multi-layers of WAs and VFOs are displayed. From the furthest to the closest to the source, the layers are labeled as $pion_{\#1}$ (green), $pion_{\#2}$ with re-emission (orchid), $pion_{\#3}$ (blue), $pion_{\#4}$ (orange), and $pion_{\#6}$ (purple). Note that the VFO absorption features are not fully captured in the zoon-in panel; the blue-shifted Fe XXVI at around 7.2 keV is visible in the main panel. The UFO component is shown in the main panel. Refer to Table \ref{table:parameters} for parameter values. }
\label{fig: dec02}
\end{figure*}

\begin{figure*}
\includegraphics[width=\textwidth]{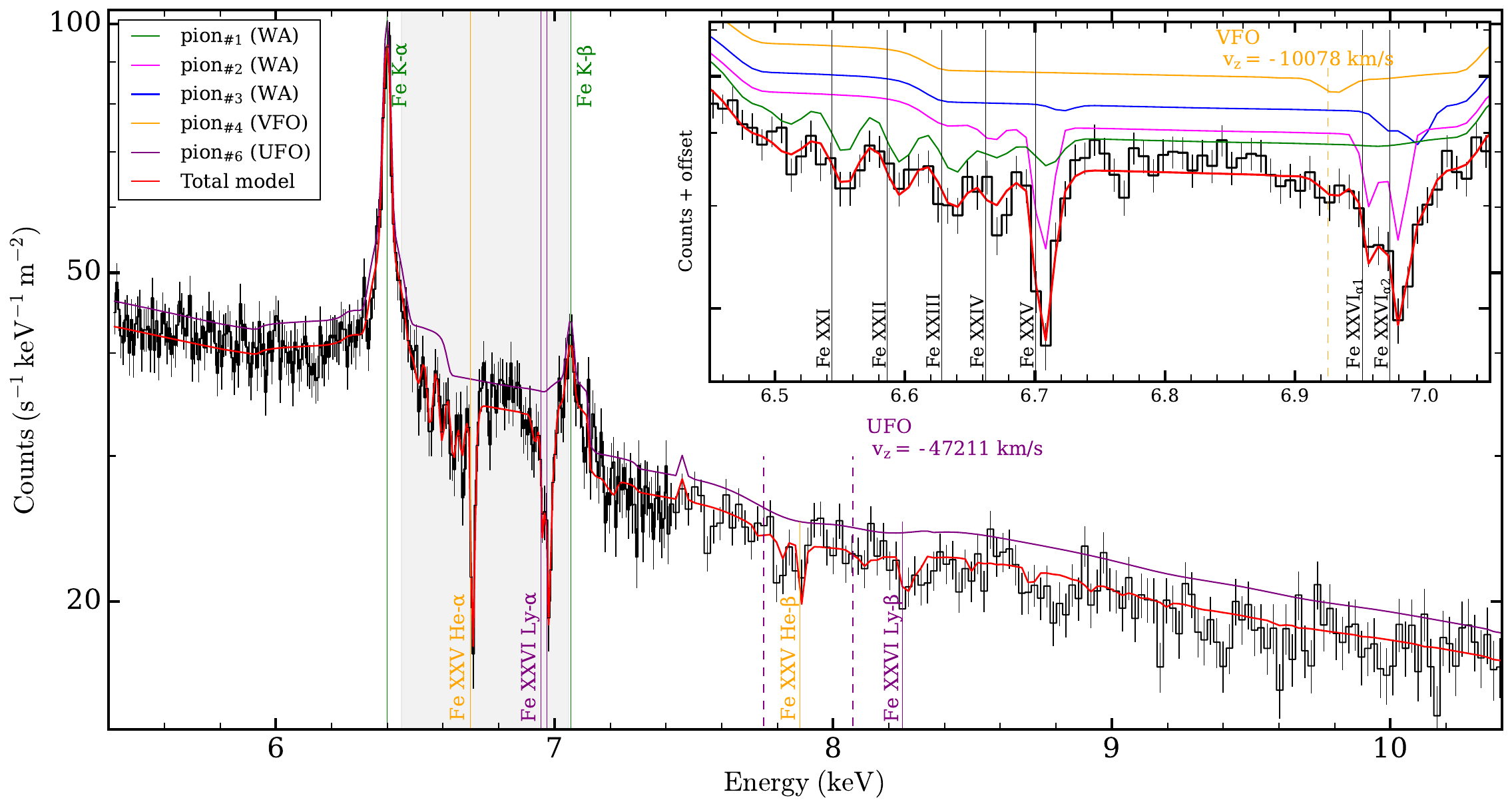}
\caption{XRISM observation of NGC 4151 on Dec 26 (Obs. 2) with the best-fit model shown as red curves. From the furthest to the cloest to the source, the layers are labeled as $pion_{\#1}$ (green), $pion_{\#2}$ (orchid), $pion_{\#3}$ (blue), $pion_{\#4}$ (orange), and $pion_{\#6}$ (purple). Compared to Obs. 1, the re-emission component was excluded due to a zero column density result in the fitting.}
\label{fig: dec26}
\end{figure*}

\begin{figure*}
\includegraphics[width=\textwidth]{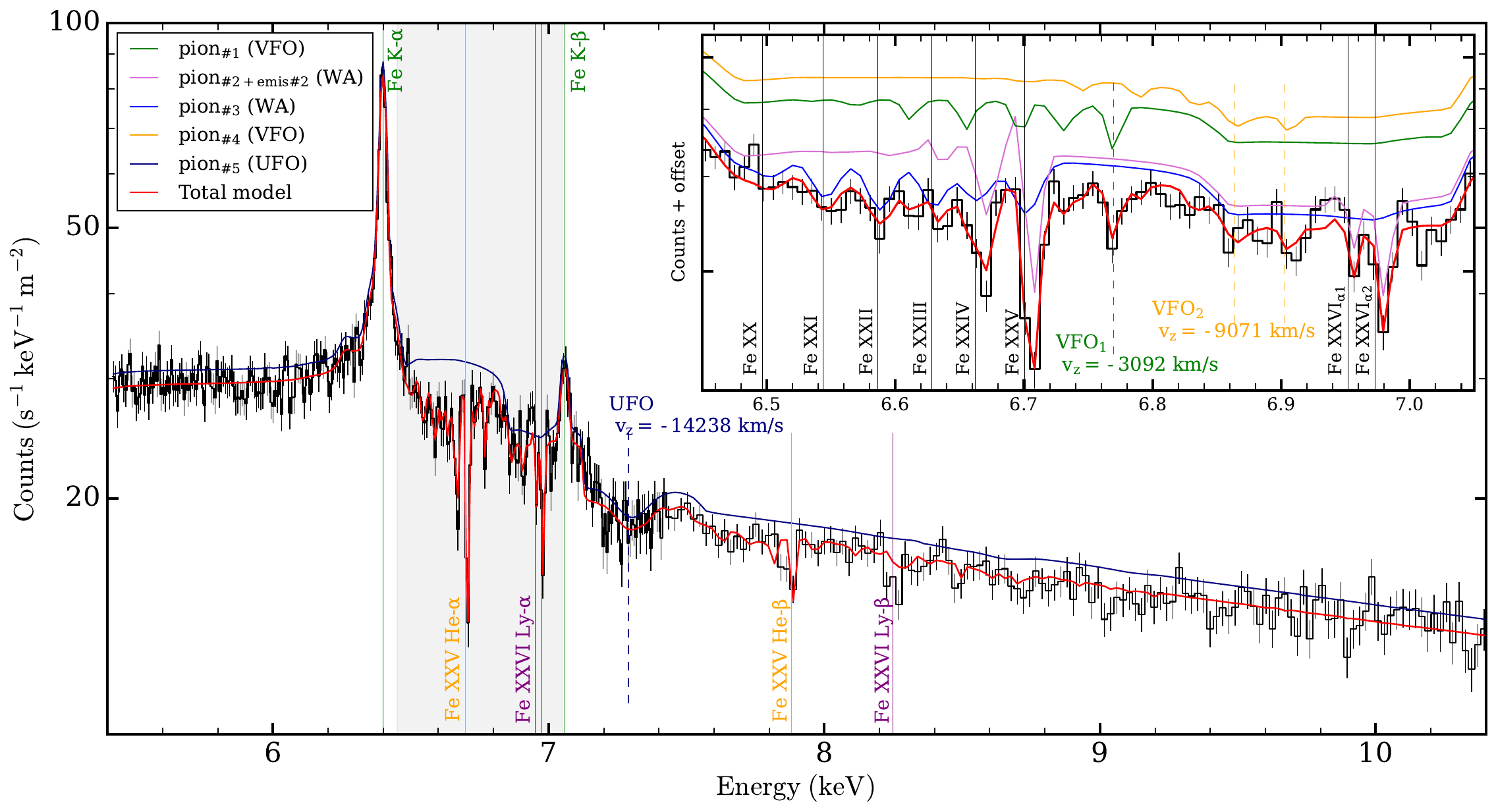}
\caption{XRISM observation of NGC 4151 on May 18 (Obs. 3) with the best-fit model shown as red curves. From the furthest to the closest to the source, the layers are labeled as $pion_{\#1}$ (green), $pion_{\#2}$ with re-emission (orchid), $pion_{\#3}$ (blue), $pion_{\#4}$ (orange), $pion_{\#5}$ (navy), and $pion_{\#6}$ (purple).}
\label{fig: may18}
\end{figure*}

\begin{figure*}
\includegraphics[width=\textwidth]{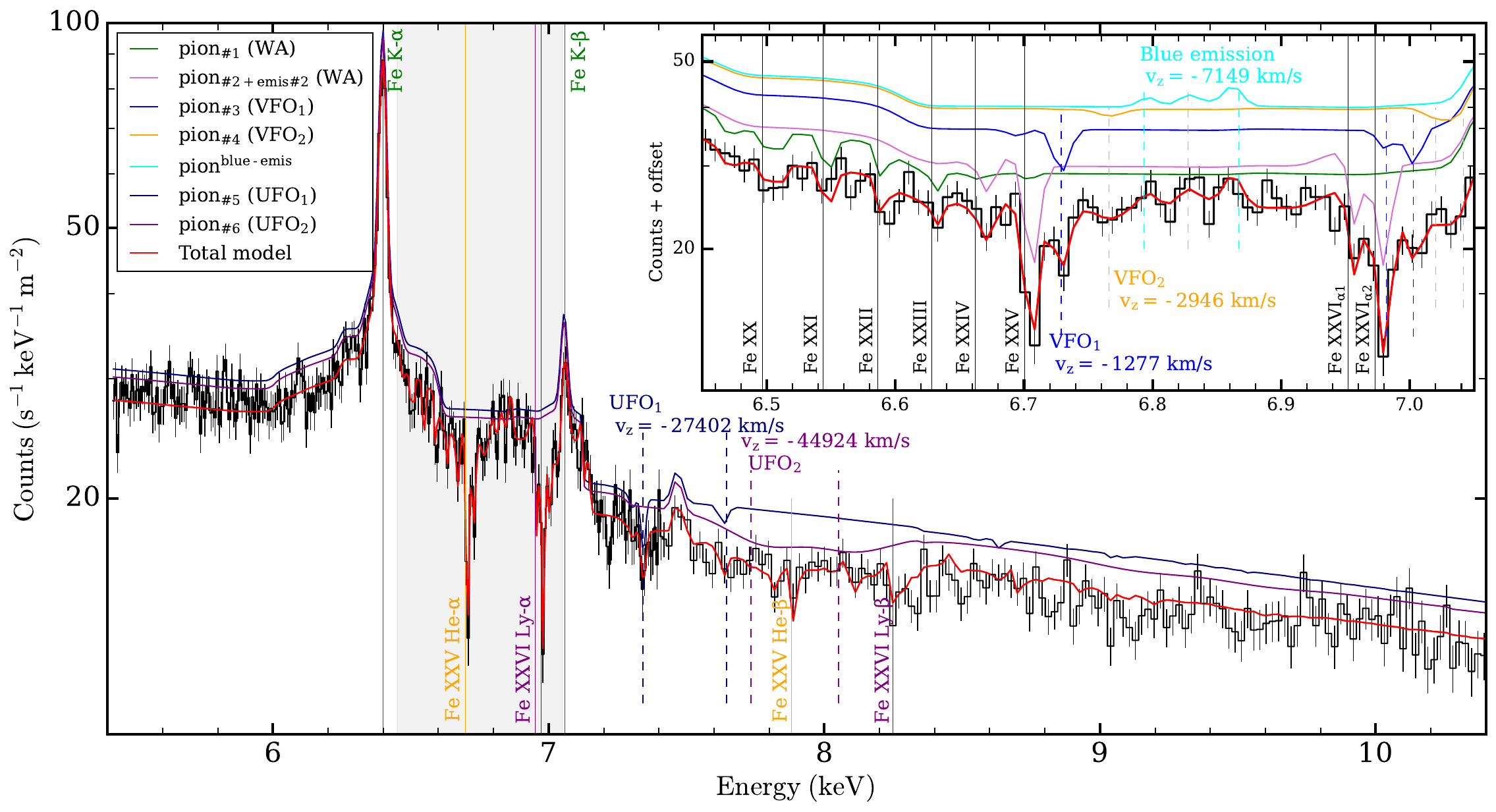}
\caption{XRISM observation of NGC 4151 on June 15 (Obs. 4) with the best-fit model shown as red curves. From the furthest to the closest to the source, the layers are labeled as $pion_{\#1}$ (green), $pion_{\#2}$ with re-emission (orchid), $pion_{\#3}$ (blue), $pion_{\#4}$ (orange), $pion_{\#5}$ (navy), and $pion_{\#6}$ (purple). The blue-shifted emission feature is shown as a cyan curve.}
\label{fig: june15}
\end{figure*}

\begin{figure*}
\includegraphics[width=\textwidth]{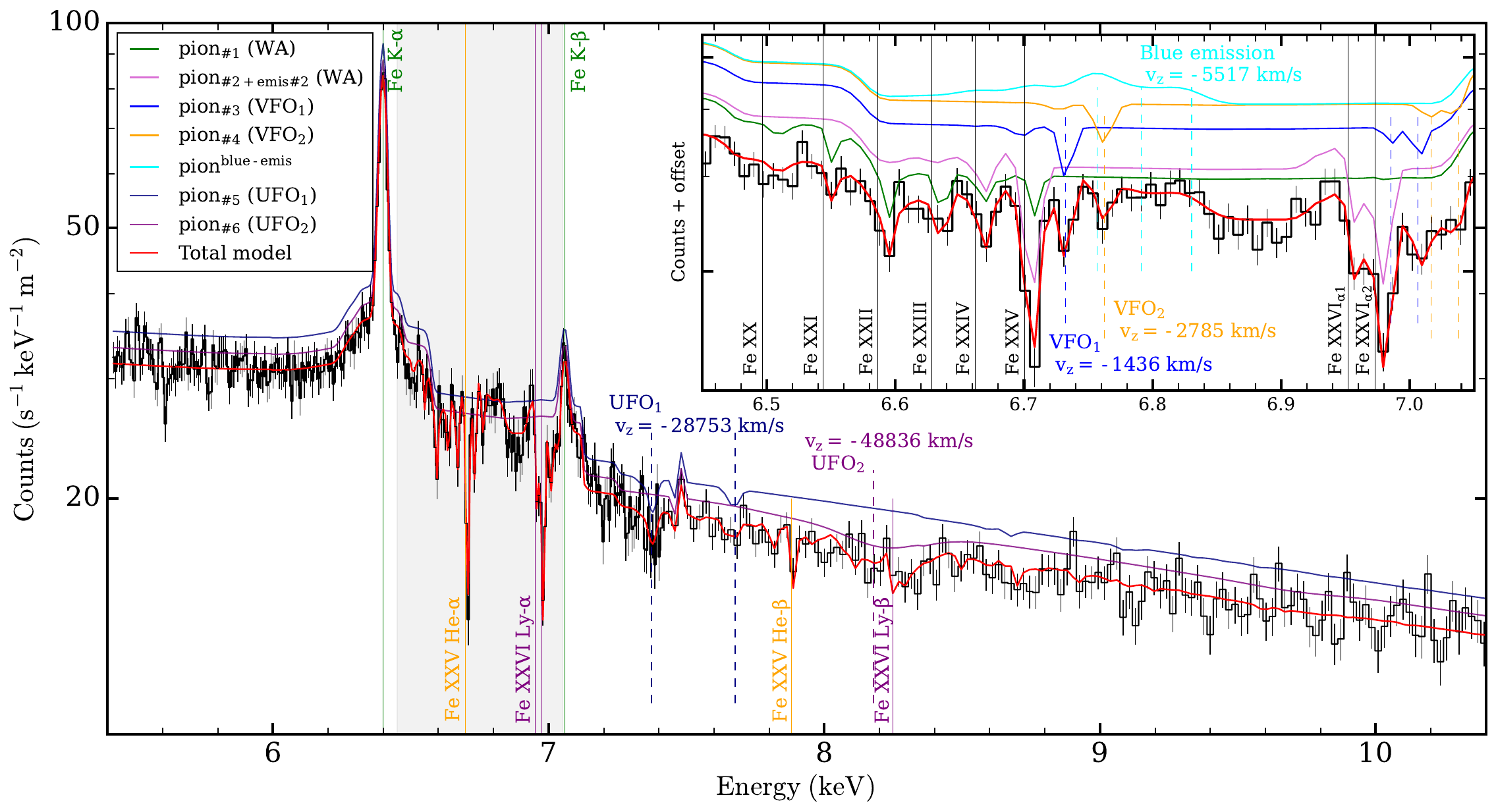}
\caption{XRISM observation of NGC 4151 on June 22 (Obs. 5) with the best-fit model shown as red curves. From the furthest to the closest to the source, the layers are labeled as $pion_{\#1}$ (green), $pion_{\#2}$ with re-emission (orchid), $pion_{\#3}$ (blue), $pion_{\#4}$ (orange), $pion_{\#5}$ (navy), and $pion_{\#6}$ (purple). The blue-shifted emission feature is shown as a cyan curve.}
\label{fig: june22}
\end{figure*}

Even with three layers of absorbers, there are still unmodeled absorption/emission features between 6.7--7.1 keV, especially for Obs. 3, Obs. 4, and Obs. 5.  To address this, an additional ``$pion_{\#4}$'' absorber was included in the model.  This component contributes to the Fe XXV and Fe XXVI line flux, which is significantly blue-shifted and differentiated from the WA line flux that is observed closer to the rest of the energy values.  Particularly in Figure \ref{fig: brd}, it is evident that these VFOs contribute absorption line flux between 6.72 -- 6.78 keV and 7.00 -- 7.03 keV, and that the flux is highest in Obs. 4, also evident in Obs. 5, but absent in Obs. 1 and Obs. 2.  This is borne out statistically: the VFO is significant and beyond $3\sigma$ threshold in Obs. 4 and Obs. 5, and is not preferred by the AIC on Obs. 1--3.   This component is also shown as the orange curve in the zoom-in panel of all five observations in Figure \ref{fig: dec02}, \ref{fig: dec26}, \ref{fig: may18}, \ref{fig: june15} and \ref{fig: june22}.   Indeed, in Obs. 4 and Obs. 5, component\#3 and component\#4 both rank as VFOs.  All four parameters of component\#4 -- $N_H$, $\xi$, $v_z$, and $\sigma_v$ -- are higher than those of component\#3, likely indicating that component\#4 traces a more extreme and energetic layer.  UFOs are known to be highly variable and we have observed modest variability in the slow WA components; potentially, VFOs are not only intermediate in velocity, but also in their variability.  This is broadly consistent with a picture wherein wind observed close to the central engine varies strongly and rapidly owing to their small size, whereas more distant components vary slowly and modestly owing to their large spatial extent.

In Figure \ref{fig: brd}, it is clear that some spectra show complexity between 6.7--6.97~keV that is indicative of narrow emission lines.  We therefore added a blue-shifted photoionized emission component to the model ($pion_{blue-emis}$) that is shaped via Gaussian broadening.  The data prefer this description over more highly ionized and red-shifted emission.  Whereas fits to Obs. 1, Obs. 2, and Obs. 3 strongly favor a column density consistent with zero, this component is highly significant in Obs. 4 and Obs. 5, with D.S $> 5\sigma$.   In those observations, the blue-shifts are very high:  $v = -7149^{+91}_{-95}~{\rm km}~{s}^{-1}$ and $v = -5494^{+240}_{-236}~{\rm km}~{\rm s}^{-1}$, respectively. They are marked as cyan curves in the Figure \ref{fig: june15} and \ref{fig: june22} in the zoom-in panels. The transient nature of this component in Obs. 4 and Obs 5, combined with an apparent absence in Obs. 1, Obs. 2, and Obs.3, demonstrates variability on a time scale of just $\sim7$~days.  The observed velocities are comparable to the free--fall velocity from the BLR, and the minimum variability time scale of $\sim7$~days is comparable to the light crossing time to the BLR.  In concert, these factors suggest that the observed emission is a failed, infalling, and potentially anisotropic component of the wind seen on the far side of the central engine.

\subsubsection{UFO components; \#5 and \#6}

To model apparent UFOs absorption features above $\sim7.1$~keV, we added two ``\texttt{$pion$}'' ($pion_{\#5}$ and $pion_{\#6}$) layers.  In these components, the turbulent velocity was allowed to vary over a broader range, of $100~{\rm km}~{\rm s}^{-1} \leq \sigma \leq 5000~{\rm km}~{\rm s}^{-1}$. The fits revealed a broad absorption component ($pion_{\#6}$, layered closest to the central engine) with a blue-shift velocity of approximately $v \sim 45000~\mathrm{km~s^{-1}} ~\mathrm{or}~0.15c$ in Obs. 1, Obs. 2, Obs. 4, and Obs. 5, and a narrower component ($pion_{\#5}$, layered outside of $pion_{\#6}$) with velocity of $v \sim 15000~\mathrm{km~s^{-1}} ~\mathrm{or}~ 0.05c$ in Obs. 3 and $v \sim 30000~\mathrm{km~s^{-1}} ~\mathrm{or}~ 0.1c$ in Obs. 4 and Obs. 5. These components are identified via Fe XXV He$\alpha$ and Fe XXVI Ly$\alpha$ absorption lines, with their relative contributions determined by the ionization parameter ($\xi$). For components with higher ionization parameters ($\log{\xi} > 3.4$), the absorption is dominated by Fe XXVI Ly$\alpha$, as seen in $pion_{\#6}$ of Obs. 1, $pion_{\#5}$ of Obs. 3, and $pion_{\#6}$ of Obs 5. these are marked with single vertical dashed lines in Figure \ref{fig: dec02}, \ref{fig: may18}, and \ref{fig: june22}. In contrast, components with lower ionization parameters show a stronger contribution from Fe XXV He$\alpha$ alongside Fe XXVI Ly$\alpha$ ($pion_{\#6}$ in Obs. 2, both $pion_{\#5}$ and $pion_{\#6}$ in Obs. 4, and $pion_{\#5}$ in Obs. 5). these are indicated by two vertical dashed lines in Figure \ref{fig: dec26}, \ref{fig: june15}, and \ref{fig: june22}. The broad ($\sigma_v > 3000~{\rm km~s^{-1}}$) absorption components significantly improve the fit, with D.S = $5.3\sigma$, $4.0\sigma$, $6.0\sigma$, and $7.6\sigma$ for Obs. 1, Obs. 2, Obs. 4, and Obs. 5 respectively. These components are shown as purple curves in the main panel of Figures \ref{fig: dec02}, \ref{fig: dec26}, \ref{fig: june15} and \ref{fig: june22}.  The secondary, narrower UFO components are not required in Obs. 1, Obs. 2 and are statistically required in Obs. 3 with D.S = $3.9\sigma$.  In the case of Obs. 1 and Obs. 2, the best-fit column density includes zero.  In Obs. 4 and Obs. 5, the secondary UFO component is marginal in terms of Gaussian equivalent sigma, and are not significantly preferred by the AIC. These features are shown as blue curves in the main panels of Figures \ref{fig: may18}, \ref{fig: june15}, and \ref{fig: june22}.

Even among the broad UFO features, there is significant variability in key parameters, especially in the column density.  In Obs. 5, the broad UFO has a column density of $1.54\pm 0.09\times 10^{23}~{\rm cm}^{-2}$, whereas the column in Obs. 2 is $N_{H} = 4.5^{+0.7}_{-0.6}\times 10^{22}~{\rm cm}^{-2}$, a difference of a factor $\sim3$.  The broad UFO components for Fe XXV and Fe XXVI lie at a range of E$=7.7-8.1$~keV in the host frame, or 7.6--8.0~keV in the observed frame across different observations.  This is broadly compatible with the tentative UFO reported by \cite{Tombesi_2010} at 7.8~keV.   At CCD resolution, a highly shifted and broad wind feature would be blended with Fe XXV He${\beta}$ and Fe XXVI Ly${\beta}$, as well as continuum flux.  The sensitivity and resolution of Resolve make it possible to clearly detect the narrow Fe XXV He${\beta}$ and Fe XXVI Ly${\beta}$ lines associated with slower components.  The difficulties of UFO searches at CCD resolution likely placed a higher floor on the column densities that can be detected; with Resolve, it is clear that UFOs are indeed strongly variable, but it is likely an effect of variable wind parameters rather than an ``on or off'' binary.  The velocity width of the broad UFOs that we have detected is unsurprising.  A velocity with of $\sigma = 5000~{\rm km}~{\rm s}^{-1}$ translates to $\Delta E =130$~eV at 8.0~keV -- less than the resolution of CCD spectrometers at that energy.  Single-bin UFOs in CCD spectra are compatible with the broad UFOs that we have detected with Resolve.

\subsection{Additional checks with NuSTAR}
The best-fit models for the Resolve data were recorded and applied to the four simultaneous NuSTAR observations in the 2-–50~keV range. The fitting was performed with only a subset of free parameters: the power-law index, normalization, and ISM absorption column density and covering factor. The lower panel of Figure \ref{fig: brd} shows the comparison spectra of XRISM and NuSTAR. The best-fit power-law indices for the NuSTAR observations are 1.65 (Obs. 2), 1.60 (Obs. 3), 1.65 (Obs. 4), and 1.64 (Obs. 5), all within 0.1 of the corresponding XRISM best-fit values. The differences compared to the XRISM best fits are -0.07, +0.02, -0.06, and -0.06, respectively. Overall, all four NuSTAR spectra exhibit no significant deviations from the XRISM spectra.

\subsection{The spectral model in full}

The total best-fit model can be summarized as follows:

\begin{multline}
     ((bb + (pow * etau_{low} * etau_{hi} + 3 \times (mytorus * spei))) \\ * line_{1} * line_{2} *  pion_{\#6} * pion_{\#5} * pion_{\#4} \\ * pion_{\#3} * pion_{\#2} * pion_{\#1} 
 \\ + vgau*pion^{emis\#2} + vgau*pion^{blue-emis})  \\ * hot * reds
\end{multline}

\begin{figure*}
\includegraphics[width=\textwidth]{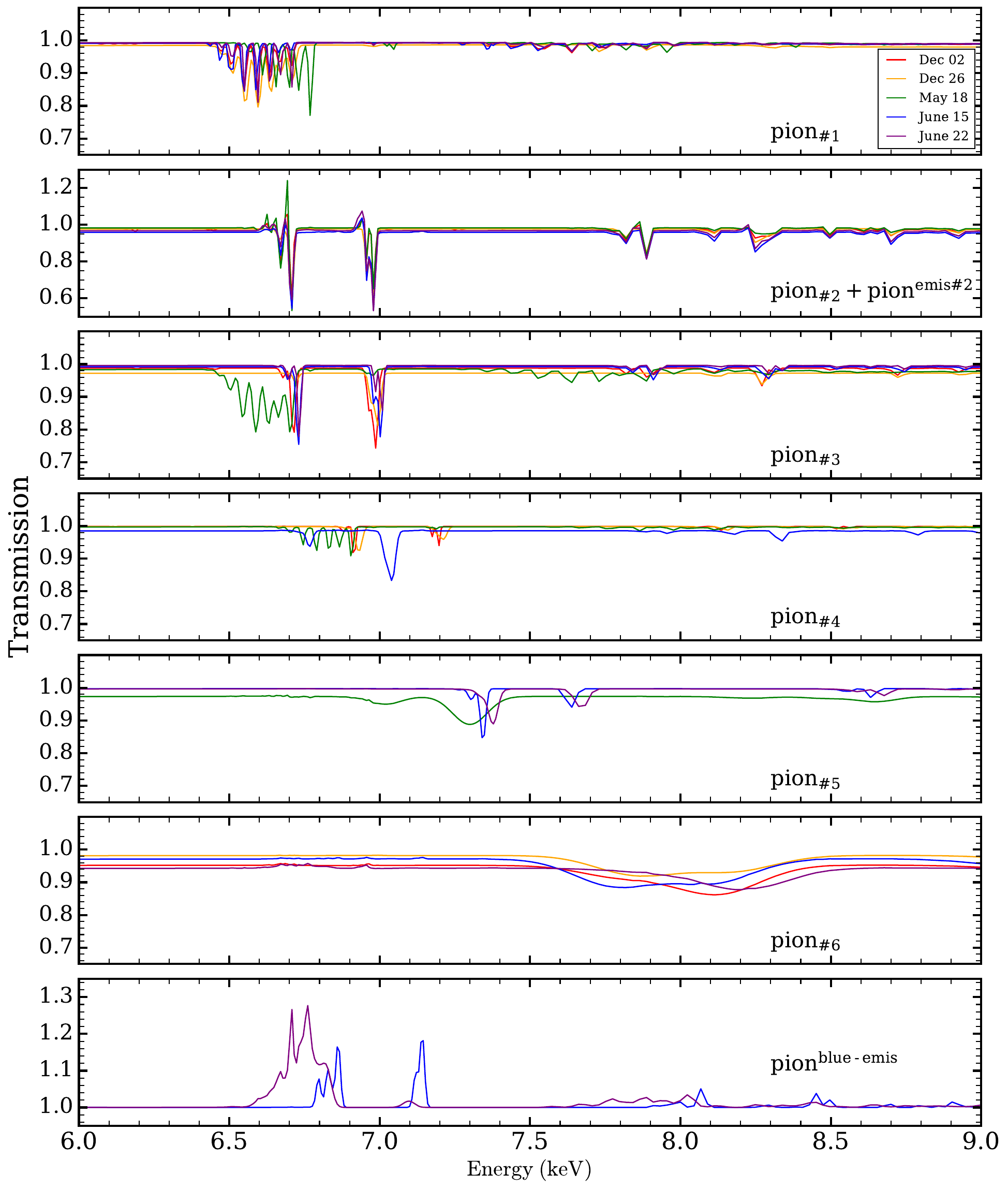}
\caption{The transmission profile of individual ``$pion$'' components across all five observations, calculated by dividing the full best-fit model by the model with the specitied component removed. Each panel corresponds to a different ``$pion$'' components, ordered from top to bottom as: $pion_{\#1}$, $pion_{\#2} + pion^{emis\#2}$, $pion_{\#3}$, $pion_{\#4}$, $pion_{\#5}$, $pion_{\#6}$, and $pion^{blue-emis}$, following the labeling in Table \ref{table:parameters}.}
\label{fig: allpion}
\end{figure*}

Through iterative fitting in both the narrow (5.4--10.4~keV) and broad (2.4--17.4~keV) energy ranges, as well as fine adjustments of parameters between fits, the total spectral model was refined to comprehensively describe the most prominent photoionized absorbers, emitters, the continuum, and fluorescence emission lines.  The best-fit parameters for all photoionization components, the $C$-value for all the observations, and the expected $C$-value (calculated automatically by SPEX) are listed in the Table \ref{table:parameters}. The transmission of each ``\texttt{$pion$}'' component is presented in Figure \ref{fig: allpion}.

\begin{figure*}
\includegraphics[width=\textwidth]{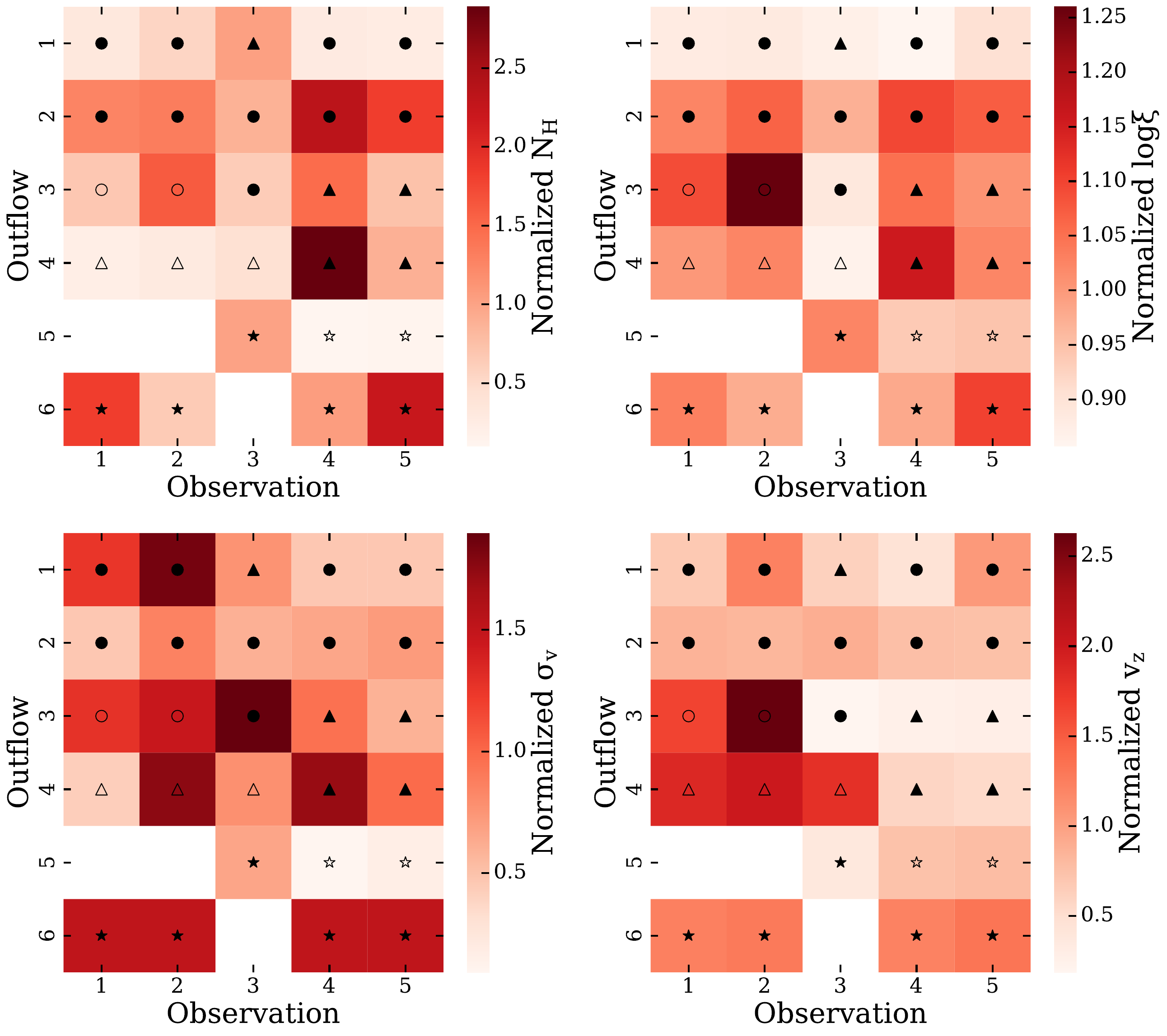}
\caption{Heapmaps shows the variability of four outflow parameters across five observations for up to six outflow components normalized differently by the mean values for each outflow type. The four panels display column densities $N_H$ (upper left), ionization parameters $\log{\xi}$ (upper right), turbulence velocities $\sigma_v$ (lower left), and velocities along our line-of-sight $v_z$ (lower right). The markers in the center represent different types: circles for WAs, triangles for VFOs, and stars for UFOs. The hollow markers indicate components with low detection significance $\sigma < 3$.}
\label{fig: heatmap}
\end{figure*}

\begin{figure}
\includegraphics[width=0.49\textwidth]{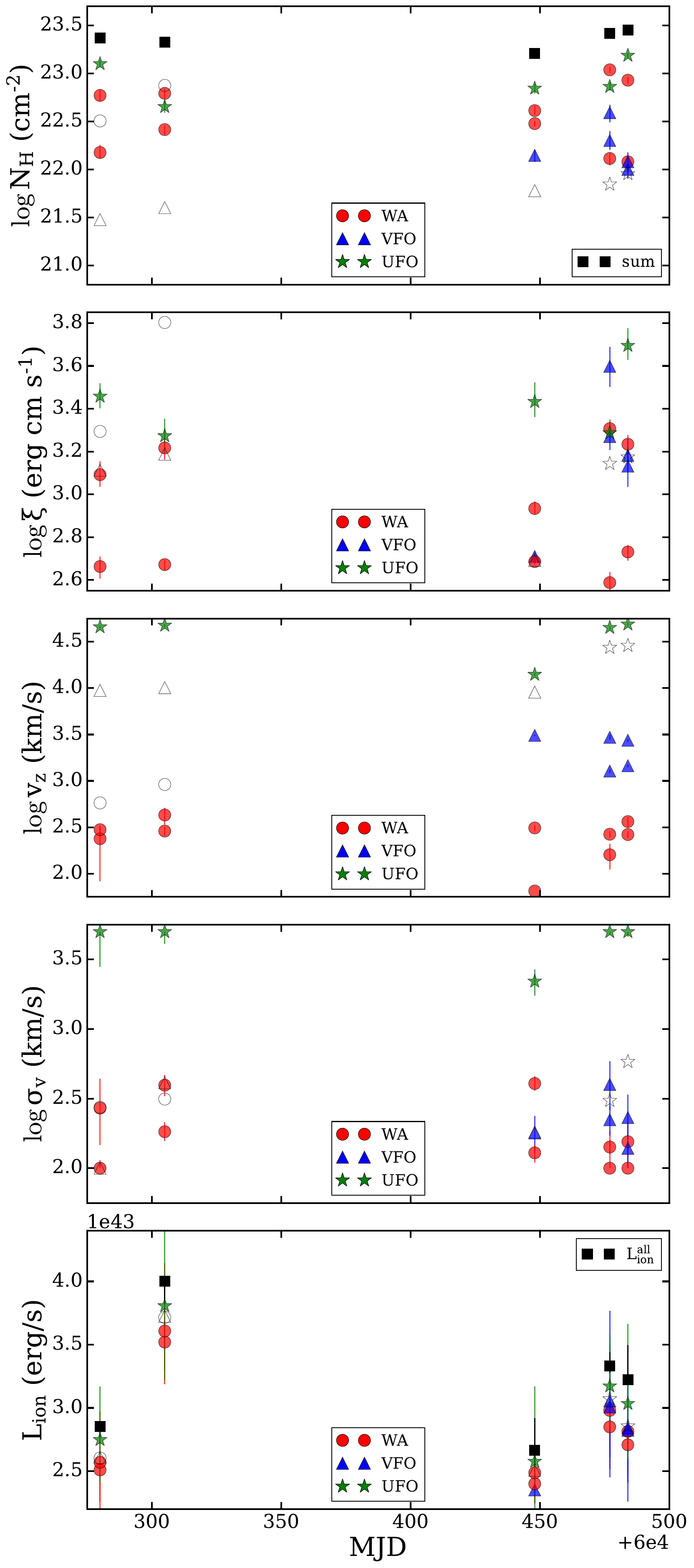}
\caption{The values of four outflow parameters and ionization luminosities as a function of time (MJD) for different outflow types (red circle for WAs, blue triangles for VFOs, green stars for UFOs). Hollow data points indicate components with low detection significance ($\sigma < 3$). The black squares in the top and bottom panels represent the sum of the column densities $N_H$ and the total ionizing luminosity for each observation. The ionization luminosity seen by each component is calculated by multiplying the \texttt{lixi} output from \texttt{pion} by its corresponding value of $\xi$.}
\label{fig: propMJD}
\end{figure}

\begin{figure}
\includegraphics[width=0.49\textwidth]{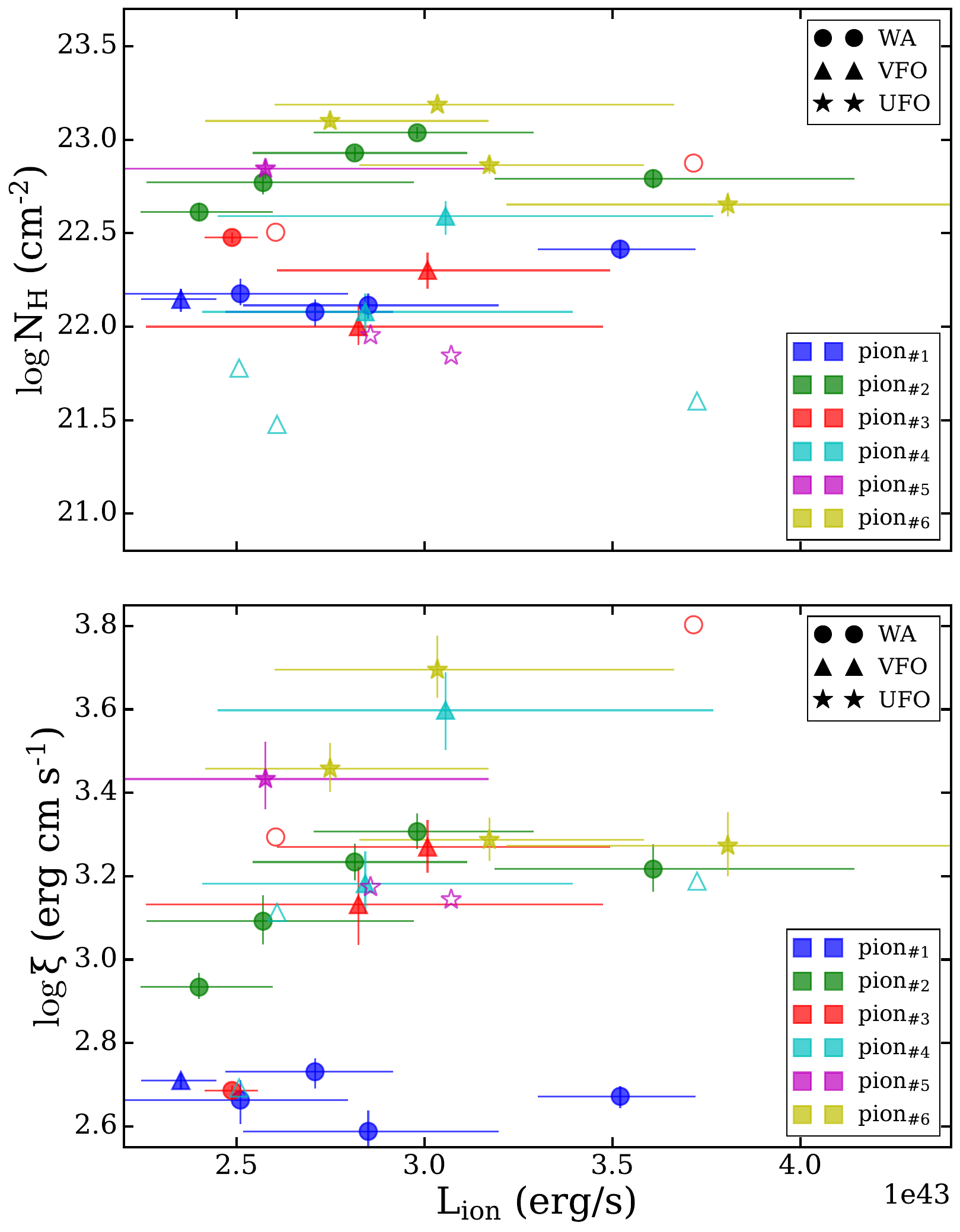}
\caption{The values of column densities (upper panel) and ionization parameters (lower panel) as a function of ionization luminosity seen by each component for different outflow types (circle for WAs, triangles for VFOs, stars for UFOs). Hollow data points indicate components with low detection significance ($\sigma < 3$). }
\label{fig: prop_Lion}
\end{figure}

Figure \ref{fig: heatmap} shows the color map of the four free wind parameters (normalized by their mean values) separately for different outflow types in each absorption component across five observations.  This is one means of tracking the relative variations of individual components over time. Figure \ref{fig: propMJD} displays the values of the four free wind parameters and the ionization luminosity ($L_{\mathrm{ion}}$) seen by each component as a function of time (MJD), categorized by outflow type.  The ionization luminosity seen by each component is calculated by multiplying the \texttt{lixi} by its corresponding value of $\xi$ (the \texttt{lixi} is defined as $L_{\mathrm{ion}} / \xi$ for each pion component, based on the ionization balance calculation across all the outflow layers in ``\texttt{$pion$}''). The observed variability in $N_H$, $\xi$, $v_z$, and $\sigma_v$, as well as the types of the component, suggest dynamic changes in the outflow structure. As shown in the top panel of Figure \ref{fig: propMJD}, the total $N_\mathrm{H}$ is neither conserved across the five observations nor conserved for different types of outflows, although they remains the same order of magnitude ($\sim 10^{23}~\mathrm{cm^{-2}}$). The total column densities of wind components in Obs. 3 ($N_\mathrm{H}^{all} = 1.61 \times 10^{23}~\mathrm{cm^{-2}}$) is noticeably lower than those in Obs. 5 ($N_\mathrm{H}^{all} = 2.82 \times 10^{23}~\mathrm{cm^{-2}}$). This suggest that gas may appear or disappear along our line of sight, in contrast to source like NGC 7469 where the total column density ($N_\mathrm{H}$) of the WAs remains conserved while individual components vary their ionization in response to changes in the ionizing continuum \citep{Mehdipour_2018}. Among the absorption components, only $pion_{\#2}$ and $pion_{\#3}$ show a potential positive correlation between their ionization parameters and the ionizing luminosity, as shown in the bottom panel of Figure \ref{fig: prop_Lion}. This may indicate that WAs are more responsive to radiative driving, or possibly located closer to the ionizing source (and hence could be associated with failed winds). For other components, we observe no clear correlation between ionization parameters and luminosity, which may either be due to lack of a measurable time delay or suggest that we are probing different regions of a spatially complex outflow.

\section{Derived wind parameters and synthesis} \label{sec: wind properties}
In this section, we use the best-fit parameters from each observation to derive key outflow characteristics that are not observed directly. These include the wind density profiles, and constraints on the launching radius, outflow rate, and kinetic power in each wind component that produces absorption features.

\subsection{The density profiles} \label{sec:AMD}
The density profile of the outflows can be estimated using the Absorption Measurements Distribution (AMD), defined as the distribution of the column density as a continuous function of the ionization parameter: $\mathrm{AMD} \equiv dN_{\mathrm{H}} / d \log{\xi} \propto \xi^m$ \citep{Behar_2009, Holczer_2007}.  If all outflow layers are part of the same wind -- if they are all produced via the same mechanism or the same ratio of mechanisms at all radii -- then the density profile, $n(r) \propto r^{\alpha}$, can be inferred from the AMD.
  
For large-scale outflows, the AMD slope, $m$, obtained by fitting $\log{N_H} = m \log{\xi} + b$ to the dataset, directly relates to the density profile exponent $\alpha$ as follows \citep{Behar_2009}:
\begin{equation} \label{eq: a_to_alpha}
    \alpha=\frac{1+2 m}{1+m} \pm \frac{\Delta m}{(1+m)^2}
\end{equation}

\begin{figure*}
\includegraphics[width=\textwidth]{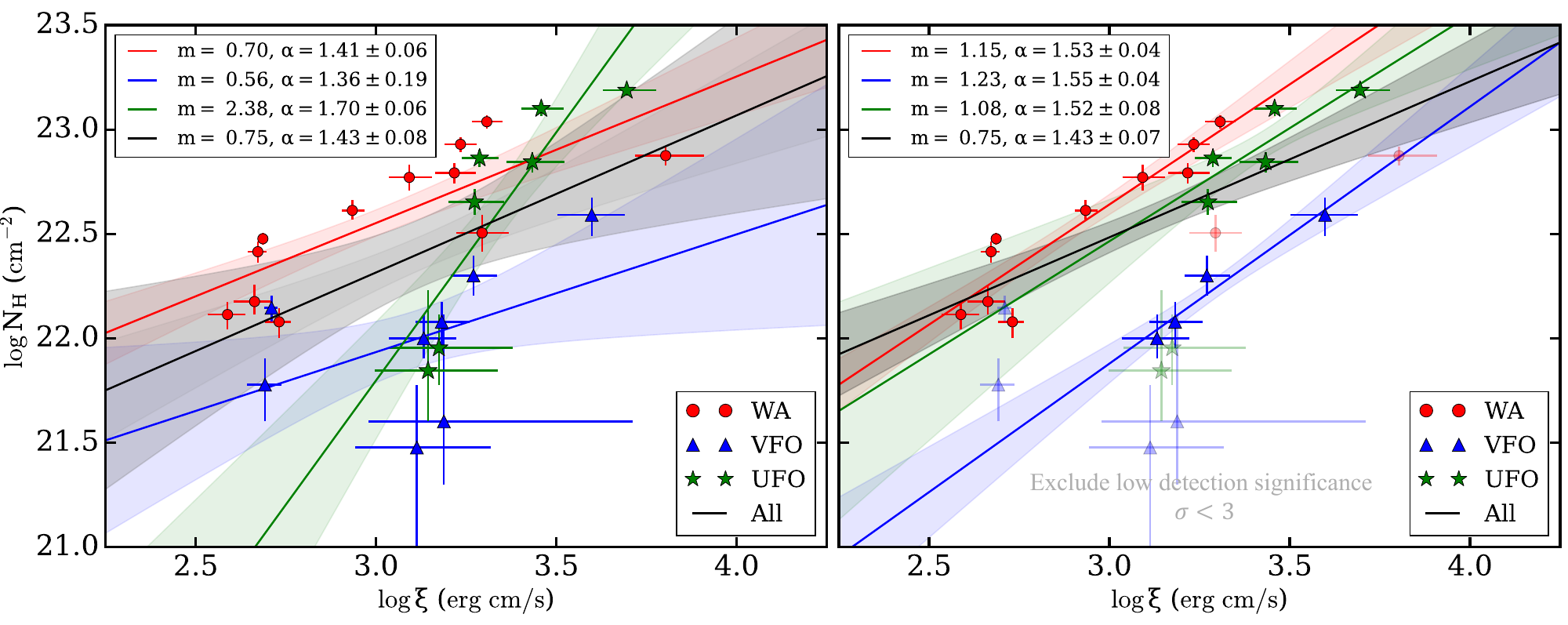}
\caption{The relation between column density $N_H$ and ionization parameter $\xi$, giving the absorption measure distribution (AMD) of the outflow components. The best-fit lines for different types of outflows (red dots for WAs, blue trangles for VFOs, and green stars for UFOs) are shown, with shaded regions indicating the $1\sigma$ confidence level. The best fit slope $m$ and the corresponding density profile index $\alpha$ with their $1\sigma$ uncertainties is presented in the legend. The right panel presents results excluding outflows with detection significance below $3\sigma$.}
\label{fig: AMD}
\end{figure*}

We used Python's \texttt{statsmodels} module to implement an Ordinary Least Squares (OLS) linear regression for the WAs, VFOs, and UFOs, both separately and collectively. Figure \ref{fig: AMD} displays the best-fit relations between column densities and ionization parameters, along with the corresponding density profile index $\alpha$ derived using Equation \ref{eq: a_to_alpha}. The results achieved after excluding the outflows with low detection significance values (below $3\sigma$) are shown in the right panel of Figure \ref{fig: AMD}.
The slope of the WAs is constrained to be $m = 0.70 \pm 0.17$, corresponding to a density profile index of $\alpha = 1.41 \pm 0.06$. For the VFOs, the slope is $m = 0.56 \pm 0.45$, indicating $\alpha = 1.36 \pm 0.19$, though large uncertainties arise from outflows with low column densities and detection significance. The slope for UFOs is steeper, mainly influenced by those outflows with low column densities and high error bars on their properties, with $m = 2.38 \pm 0.64$ corresponding to $\alpha = 1.70 \pm 0.06$. 

The WAs and VFOs exhibit slopes consistent with the predictions of Blandford-Payne (BP) winds ($\alpha = 1.5$; \citealt{Blandford_Payne_1982}) within error bars. The slope derived by fitting all potential UFOs deviates significantly from BP wind, instead falling between radiatively driven winds ($\alpha = 2$; \citealt{Kazanas_2012}) and BP winds.  However, when excluding low-significance UFOs -- these are the slower, narrower, and less ionized cases detected in Obs. 4 and Obs. 5 -- the slope becomes flatter and aligns with BP winds.  These two low-ionization UFOs may represent discrete constituents of different UFOs originating from a different region of the disk, where shielding by inner UFOs preserves line-driving efficiency \citep{Proga_2000, Proga_Kallman_2004}. 
Similarly, excluding low-significance VFOs detected in Obs. 1, Obs. 2, and Obs. 3 - which happen to be the highest velocity cases - improves the slope constraint, making it consistent with BP winds. Likewise, removing low-significance WAs, which happen to have the highest-velocity inner WAs in Obs. 1 and Obs. 2, enhances linear fit bringing the slope into agreement with BP winds.   When all outflow types are analyzed together, the results remain consistent with BP winds, regardless of whether low-detection outflows are excluded.

These high-velocity low-significance VFOs tend to steepen the AMD slope, reminiscent of the falloff of the $N_H -\xi$ relation seen in Figure 13 of \citet{Waters_2021}, where a clumpy, thermally unstable state develops. The similarity between our AMD and those from the simulation suggests that the observed WAs are tracing a later, clumpy stage of thermally driven wind. The steep drop around $\log{\xi} \sim 3.2$ for these VFOs with high velocities is suggestive of the early stage for the formation of clumps.

\subsection{Launching Radius and Volume Filling Factor} \label{sec: r_fv}

Assuming that the radial size of an absorbing wind ($\Delta r$) does not exceed the radius at which it is launched, $N_H = n \Delta r \leq f_v n r$, where $f_{v}$ is the volume filling factor along the flow ($f_{v} = 1$ for a homogeneous and volumetrically thick outflow). Using the definition for the ionization parameter $\xi = L_{\mathrm{ion}}/nr^2$, the upper limit for the launching radius:
\begin{equation} \label{eq: r_max}
    r_{\mathrm{max}} =  \frac{L_\mathrm{ion} f_v}{\xi N_\mathrm{H}} = r_1 f_v,
\end{equation}
where $r_1 = L_{\mathrm{ion}} / \xi N_H$ is directly calculated for each outflow layer using the best-fit values of $N_H$ and the output parameter \texttt{lixi} from ``\texttt{$pion$}'' in SPEX.

If each component is observed close to its point of origin, a minimum wind launching radius can be estimated by equating the outflow velocity to the local escape velocity.
\begin{equation} \label{eq: r_min}
    r_{\mathrm{min}} = \frac{GM}{v^2_{\mathrm{out}}} = r_2 \cos^2{\theta},
\end{equation}
where $r_2 = GM/v_z^2$ is calculated directly for each outflow layer using the best-fit values of $v_z$. The angle $\theta$ represents the inclination of the outflow stream relative to the line of sight (LOS). The outflow velocity is related to the observed projected velocity by $v_{\rm out} \cos{\theta} = v_z$.  The minimum possible values of the volume filling factor can be estimated as $f_{v_{min}} =  r_2 / r_1 \cos^2{\theta}$, which follows from the condition $r_{max} \geq r_{min}$.

\begin{figure*}
\includegraphics[width=\textwidth]{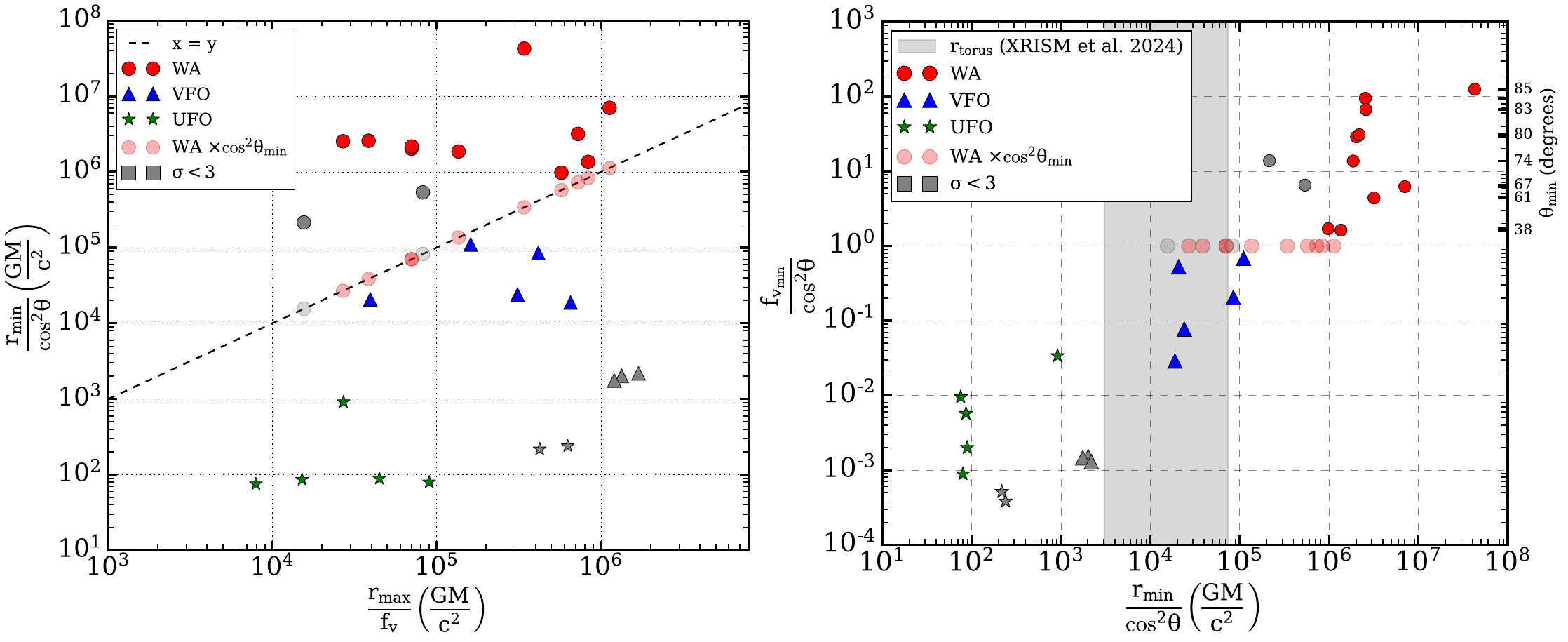}
\caption{LEFT PANEL: Comparison of the estimated minimum launching radius ($r_{min}$ in equation \ref{eq: r_min}, derived from the escape velocity condition) and the maximum launching radius ($r_{max}$ in equation \ref{eq: r_max}, constrained by the ionization parameter) for each outflow layer. The dashed line represents the equality line ($r_{min} = r_{max}$). RIGHT PANEL: Minimum volume filling factor $f_{v_{min}}$ as a function of $r_{min}$. The shaded region represents the estimated radius for the inner wall of the torus from \citealt{XRISM_NGC4151_2024}. The red circles represent warm absorbers (WAs), blue triangles correspond to Very Fast Outflows (VFOs), and green stars indicate Ultra-fast Outflows (UFOs). The transparent red circles show corrected WA values, incorporating the minimum required inclination angle to bring $f_v$ below unity. The corresponding minimum angles are plotted on the secondary y-axis in the right panel. The grey data points represent outflows with low detection significance $\sigma < 3$}
\label{fig: r_fv}
\end{figure*}

The left panel of Figure \ref{fig: r_fv} compares the estimated minimum launching radius with the maximum launching radius, without correcting for projection effects or the volume filling factors. The right panel shows the minimum volume filling factor, $f_{v_{min}}(\theta = 0)$, also without considering angle differences between the outflow stream and LOS. The shaded region represents the estimated inner wall of the torus from \cite{XRISM_NGC4151_2024}. The trend of $f_{v_{min}}$ indicates that faster outflows tend to be more clumpy (lower $f_v$). The lowest lower limit for volume filling factor is found in the fastest UFO, reaching as low as $5\times10^{-4}$. The lower limit of the volume filling factor overlaps with the radio constraints on the upper limit for volume filling factors, which range from $10^{-4} - 0.5$ \citep{Blustin_Fabian_2009}. 

For the WA wind components, the calculated minimum radius exceeds the maximum radius, yielding volume-filling factors greater than unity.  This could reflect that warm absorbers are not observed close to their point of origin.  Rather, they might be observed far from their launching point, where parameters including velocity, density, and clumping factor have changed relative to initial values.   We observe only modest changes in ``\texttt{$pion$}'' components \#1--3 between Obs. 4 and Obs. 5, which are separated by only a week -- comparable to the light travel time to the optical BLR \citep{Bentz_2006}.  However, the slow wind components appear to be significantly different in Obs. 3 and Obs. 4, separated by 28 days.  If the inner edge of the BLR lies at $r \simeq 3\times 10^{3}~GM/c^{2}$ \citep{XRISM_NGC4151_2024}, then this interval corresponds to $r \simeq 1.2\times 10^{4}~GM/c^{2}$, orders of magnitude below the radii implied by associating observed speeds with a local escape speed.   Alternatively, WA components may represent failed winds, launched with velocities that are below the local escape speed.  The radii implied by apparent variability time scales have much higher escape velocities ($v \simeq 4000~{\rm km}~{\rm s}^{-1}$) than we measure in WA components, even after accounting for plausible projection effects.  Moreover, we have detected blue-shifted gas in emission in some observations, potentially consistent with failed winds seen on the far side of the central engine.  

Alternatively, it is also possible that we view slow WA wind components at angles that differ from faster wind components launched at smaller radii. To test this, we applied a correction by assuming that a larger inclination angle between the gas stream and the line of sight (LOS) would compensate for the overestimated volume filling factor. The transparent data points for WAs in both panel of Figure \ref{fig: r_fv} represent the corrected volume filling factors and minimum launching radii, accounting for the minimum required angle adjustment that brings down the volume filling factor to unity. The corresponding minimum angles are plotted on the secondary y-axis in the right panel, which shows that the WAs launched at larger radii require larger angle differences. These winds may be concave upward relative to the accretion disk, or/and the opening angle of the torus (or outer disk) may increase the base angle of the wind launching site.  This may be consistent with the ``bowl geometry'' that is inferred in some recent optical studies of AGN, including NGC 4151 (see below).

\subsection{Mass Outflow Rate and Energetics} \label{sec: Mdot_KE}
The mass of the outflowing gas can be expressed as $\dot{M}_{\mathrm{out}} = \rho dV/dt = \mu m_p n f_v 4\pi f_{\mathrm{cov}} r^2v_{\mathrm{out}}$, where $\mu = 1.23$ is the mean atomic weight, $m_p$ is the proton mass, $n$ is the number density of the outflowing gas, and $r$ is the radial distance of the gas from the center SMBH. The covering factor is $f_{cov} = \Omega/4\pi$ ($\Omega \in [0, 4\pi]$), where $f_{\mathrm{cov}} = 1$ represents a spherical shell of gas. We assume a covering factor of 0.5 for the absorbers. The impact of different LOS angles and the inclination of the wind is not explicitly accounted for since the outflow rate is expected to vary by a factor of $\leq 2$ for reasonable angles \citep{Krongold_2005}. 

Using the definition of the ionization parameter, the mass outflow rate can be written as:
\begin{equation} \label{M_dot}
    \dot M_{\mathrm{out}} = 4 \pi f_{\mathrm{cov}}\mu m_p \frac{L_{\mathrm{ion}}}{\xi}v_{\mathrm{out}}f_v
\end{equation}
In our calculations for the values of $L_{\mathrm{ion}}/\xi$, we use the output parameters \texttt{lixi} from ``\texttt{$pion$}'' in SPEX. 

Assuming the gas has reached a terminal velocity without significant acceleration, the kinetic power (or kinetic luminosity) is given by:
\begin{equation} \label{L_ke}
    \dot E_k = \frac{1}{2} \dot M_{\mathrm{out}} v_{\mathrm{out}}^2
\end{equation}
While this assumption is valid for most disk winds that reach an approximately constant velocity after an initial acceleration phase \citep{Proga_2000, Fukumura_2010}, it may not be valid for failed winds, where gas is still accelerating or decelerating, or for outflows interacting with the surrounding medium, which may slow them down over large scales, or for the cases where winds are detected near their launching base, meaning they are not yet at their final speed.

In our five XRISM observations, the ionizing luminosities estimated in SPEX using the best-fit models are listed in Table \ref{table:observationlog}. Assuming $L_{bol} = 2L_{\mathrm{ion}}^{all}$, as appropriate for the default AGN SED estimated in CLOUDY \citep{Panda_2022, Ferland_2017}, our estimated bolometric luminosities range from $L_{bol} = 5.3\times 10^{43}$ to $L_{bol} = 8.0 \times 10^{43}$ erg/s, yielding an Eddington ratio of $\lambda_{Edd} = L_{bol} / L_{Edd} = 0.012 - 0.019$. For comparison, \cite{Kraemer_2020} estimated the bolometric luminosity of NGC 4151 as $1.4 \times 10^{44}$ erg/s by scaling average UV flux observed by HST over the past two decades with the CLOUDY standard SED. Our estimates fall within the expected variation from previous studies \citep{Kraemer_2020, Mahmoud_Done_2020}. The discrepancy may be attributed to variations in SED or intrinsic AGN variability. 

\begin{figure*}
\includegraphics[width=\textwidth]{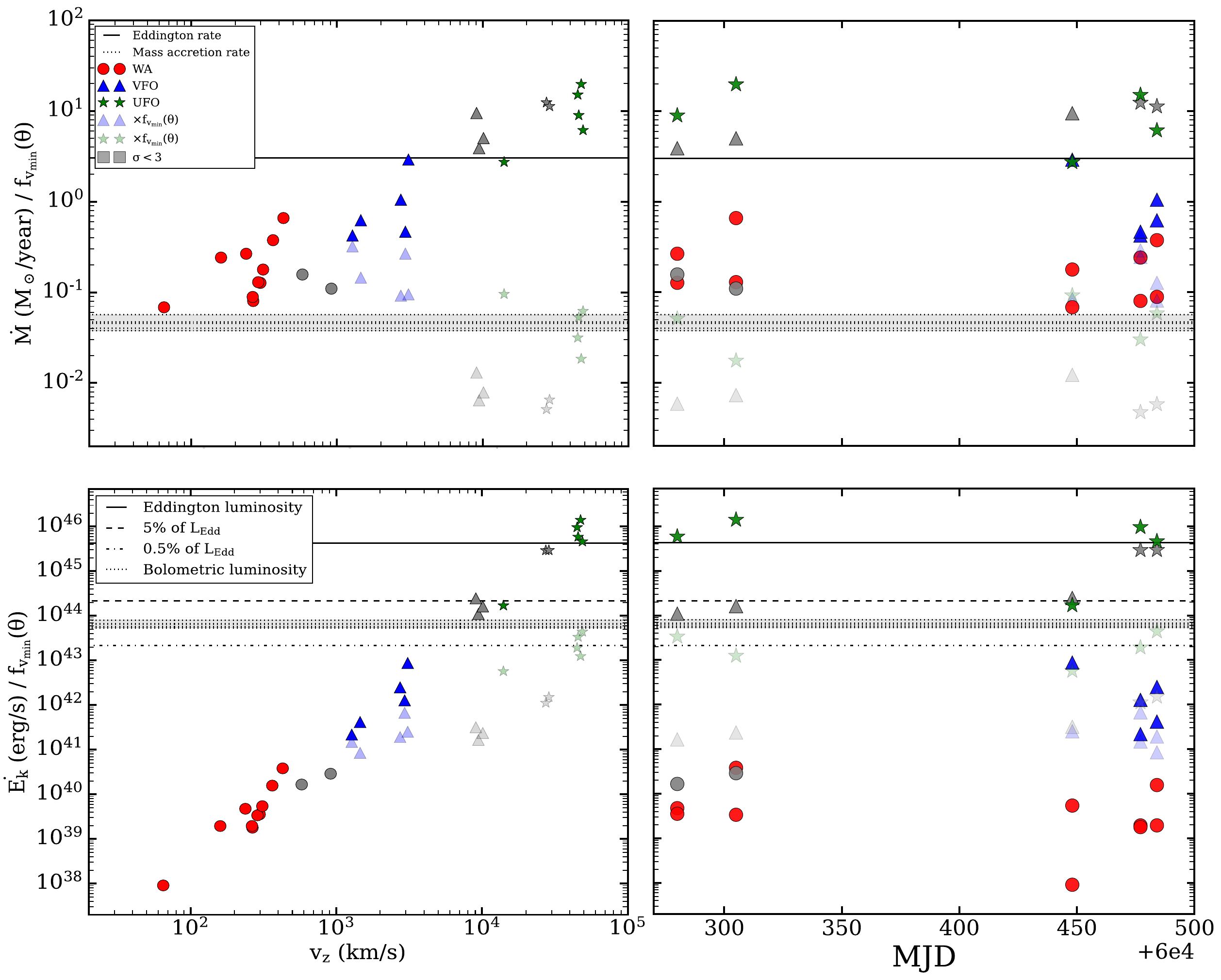}
\caption{UPPER LEFT PANEL: Mass Outflow rate per volume filling factor ($\dot{M}$) as a function of radial velocity ($v_z$) along the LOS.  Because of the large uncertainty in the volume filling factor compared with propagated statistical uncertainties from the outflow parameters, we do not include the error bar.  The solid horizontal line represents the Eddington accretion rate ($\dot{M}_{Edd}$). The dotted lines in the shaded region represent the Mass accretion rate ($\dot{M}_{acc}$) for five observations. The red circles, blue triangles, and green stars correspond to WAs, VFOs, and UFOs, respectively. The transparent markers represent values accounting for the minimum volume filling factor for VFOs and UFOs estimated in the right panel of Figure \ref{fig: r_fv}. The grey data points represent outflows with low detection significance $\sigma < 3$. LOWER LEFT PANEL: Kinetic power ($\dot{E}_k$) as a function of radial velocity. The solid, dashed, and dotted horizontal lines mark the Eddington luminosity ($L_{Edd}$), $5\%$, and $0.5\%$ of the Eddington luminosity, respectively. The dotted lines in the shaded region represent the bolometric luminosities ($L_{bol}$). Most UFOs and VFOs exceed the $0.5\%$ threshold of $L_{Edd}$.}
\label{fig: M_dotE_k}
\end{figure*}

Figure \ref{fig: M_dotE_k} shows the mass outflow rate (upper panel), and the kinetic power (lower panel) as a function of measured radial velocity ($v_z$) along the LOS (left panel) and as a function of time (MJD) in the right panel.  We do not apply the volume filling factor for the WAs due to physically inconsistent results discussed above \S\ref{sec: Mdot_KE}.  The Eddington mass accretion rate is calculated by $\dot M_\mathrm{Edd} \equiv L_\mathrm{Edd}/ \eta c^2 = 3.04~M_\odot / \rm yr$, where $L_\mathrm{Edd} \equiv 4\pi GM_{BH}m_pc/\sigma_T \simeq 4.28\times 10^{45}~erg/s$ is the Eddington luminosity for NGC 4151.  We assume a radiative accretion efficiency of $\eta = 0.4$ for rapidly spinning SMBH, probably the case in NGC 4151 \citep{Keck_2015}.   The mass accretion rate of the disk is calculated by $\dot{M}_{acc} = L_{bol}/\eta c^2$.  
For unity filling factors, the mass outflow rate increases with the outflow velocity, spanning two orders of magnitude from $0.07 - 19.8 ~M_\odot / \rm yr$, corresponding to $2\% - 6\times$ the Eddington rate and $1-525 \times$ the mass accretion rate.  All UFOs have nominal outflow rates at or above the Eddington accretion rate, suggesting that their volume filling factors must be extremely low, and/or the UFO phase is ephemeral.  Indeed, for our lowest estimates of the filling factor, all wind components may expel far less than the Eddington mass accretion rate.  However, even the lowest outflow rate is still $\sim 0.2\%$ of the Eddington mass accretion rate, and most of the outflow rates still exceed the mass accretion rate. This suggests that the winds play a significant role in transporting mass to the surrounding environment.

The lower panel of Figure \ref{fig: M_dotE_k} shows that the kinetic power of most outflows is below the Eddington luminosity, except for the broad UFOs for unity volume filling factors.  The kinetic power fraction relative to Eddington luminosity ranges from $\dot{E_k}/L_{Edd} = 2\times 10^{-8}$ to 3, spans $2\times10^{-8} - 1\times10^{-6}$ for WAs, $5\times10^{-5}-6\times10^{-2}$ for VFOs, and $4\times10^{-2} - 3$ for UFOs.  All UFOs and most of the VFOs have kinetic powers that exceed $0.5\%$ of the Eddington luminosity, the minimum energy required to sufficiently initiate the feedback and influence SMBH growth and host galaxy evolution \citep{Di_2005, Hopkins_2010}. UFOs have a strong potential to exceed $5\%$ of the Eddington luminosity if the column filling factor is bigger than $\sim 0.01$. Even applying our estimate for the minimum volume filling factor, two UFOs exceed $0.5\% L_{Edd}$.  For conservative assumptions, then, the VFOs and UFOs that are observed may represent strong feedback that meets the threshold to halt star formation in the galactic bulge. In contrast, WAs do not have the kinetic power to alter the host galaxy, but they likely play a key role in the accretion flow by removing or cycling a substantial fraction of the mass accretion rate.

\subsection{Can reflection explain the UFOs?}

To explore whether the broad UFO absorption features observed around 8 keV could instead be disk reflection, we conducted an alternative fit for Obs 4.  In this model, all the UFO components were removed and the narrowest ``\texttt{$mytorus$}'' component was replaced with the standard ``\texttt{$xillver$}'' \citep{Garcia_2013}, using the table model (\texttt{xillver-a-Ec5.fits}) converted into SPEX compatible format.  This component was then blurred using  ``\texttt{$spei$}'', with the same configuration as in our main model.

We fixed the black hole spin to $a = 0.7$, the outer disk radius at a fiducial value of 10,000 $GM/c^2$, and assumed an Euclidean emissivity of $q = 3$.  The inclination of the ``\texttt{$xillver$}'' and ``\texttt{$spei$}'' was linked, and the power-law index was tied to the underlying continuum.  The power-law energy cutoff $E_{\mathrm{cut}}$ and the iron abundance $A_{\mathrm{Fe}}$ were fixed at $300$ keV and solar abundance.  
The redshift parameter was fixed at zero (the separate ``\texttt{$reds$}'' component accounts for the redshift).  We explored two versions of the model: one with neutral reflection ($\xi = 0$) and another with mildly ionized reflection ($\xi = 200$).  The latter case is the ionization at which reflection models predict the strongest ionized Fe~K absorption lines in addition to much stronger emission lines (see \citealt{Garcia_2013}).
The remaining free parameters were the inner radius $r_{in}$, inclination $i$, and the normalization $Norm$.  We performed the same iteration of fitting between broad (2.4--17.4 keV) and narrow (5.4--10.4 keV) bands as before. The power-law index and norm are left free during the broadband fit.  All the other parameters are fixed at their previous best-fit values.  

Neither the neutral nor the ionized reflection model could satisfactorily explain the data. The best-fit statistics for the two different models with $\xi = 0$ and $\xi = 200$ in ``\texttt{$xillver$}'' are $C/d.o.f = 980/399$ and $C/d.o.f = 1598/399$ and are shown as green and blue curves in Figure \ref{fig: june15xillver}.  Both models left strong residuals near $\sim 8~keV$, where the UFO absorption was prominent.  An ionized relativistic reflection would produce a small dip around $\sim 8~keV$ but a strong emission around $\sim 6.5~keV$ that is not present in the data. 

The simultaneous NuSTAR observation, which extends spectral coverage to 50 keV as described in Section \S\ref{sec: Nustar}, when fitted with best-fit XRISM models, shows no prominent curvature or excess flux above 10 keV that would visually suggest a strong reflection hump. This suggests that the high-energy continuum is broadly consistent with a power-law without obvious reflection features. However, this test is limited in that no formal reflection component was fitted to the NuSTAR spectra. Thus, while our results likely rules out reflection as the origin of the specific UFO absorption feature, we do not exclude the possibility that a much broader relativistic reflection component from the inner disk, acting over a wide band (e.g., $> 20$ keV), may still be required -- though modeling that component lies beyond the scope of this work and will be study in a future work (R. Boissay-Malaquin et al. 2025, in preparation). 

An alternative scenario proposed in previous studies is that some of the broad absorption features attributed to UFOs may instead arise from resonance absorption of the reflected spectrum in a layer of ionized material at the surface and corotate with the disk \citep{Gallo_Fabian_2011}. Such models have been shown to reproduce some UFO-like features without invoking high-velocity outflows \citep{Gallo_Fabian_2013}. Implementing this scenario would involve convolving a photoionized absorber (e.g. ``\texttt{$pion$}'') with the same relativistic blurring applied to the reflection component. While this approach is promising, we defer a full exploration of this absorption-on-reflection model to future work.

\section{Discussion} \label{sec:discussion}
We have analyzed the five XRISM PV observations of the Seyfert 1.5 AGN NGC 4151, revealing rich complexities and multiple wind layers that span orders of magnitude in key properties, including projected velocity.  These outflow layers include different combinations of Warm Absorbers, Very Fast Outflow and Ultra Fast Outflows.  In different observations, variations in these components produce highly variable absorption and re-emission spectra. The density profiles, launching radii, volume filling factors, and energetics were estimated for all of the outflow components.  The fastest outflows may provide strong feedback that influences the host galaxy. The derived density profiles follow $n(r) \propto r^{-1.5}$ (in Section \S\ref{sec:AMD}),  consistent with magnetocentrifugal winds as per \cite{Blandford_Payne_1982}. However, while magnetic driving likely plays a key role, thermal pressure and radiation pressure may contribute to driving specific components of the total X-ray outflow.  A broad examination of the derived density profiles, constraints on launching radii and filling factors, and the facts of blue-shifted wind components in emission and likely failed winds in absorption suggests a complex outflow with significant clumping, likely axial asymmetry, and strong variability.  In this section, we discuss the implications of these results for wind launching mechanisms, and aim to form an improved view of the structure of the accretion-driven disk wind in NGC 4151.  We also address potential caveats and limitations of the analysis, and propose directions for future work. 

\subsection{Wind launching mechanisms} \label{sec: Launching Mechanisms}
Thermal pressure can launch gas at radii greater than $0.1\times R_{IC}$ \citep{Begelman_1983, Woods_1996}, or radii as small as from 0.01$\times R_{IC}$ \citep{Proga_Kallman_2002} when electron scattering pressure contributes close to the Eddington limit, where $R_{IC}$ is the Compton Radius where the escape velocity equals the isothermal sound speed at the Compton temperature.   Given the measured thermal disk temperature of $kT \sim 33.6~eV$ at peak emissivity (see \S\ref{sec: continuum modeling}), the Compton radius is $R_{IC} \sim 1.7 \times 10^7~GM/c^2$. The launching radius of all outflows calculated in both methods (see \S\ref{sec: r_fv} and Figure \ref{fig: r_fv}) are within 0.1$\times R_{IC}$.  Some WA components could be in between 0.01 and 0.1 $R_{IC}$ (or $\sim 10^5 -10^6~GM/c^2$), but we have observed NGC 4151 well below the Eddington limit.  Moreover, the efficiency of the luminosity in generating a wind via Compton heating is quantified by the parameter $L_{bol}/L_{cr}$ (where $L_{cr} = 0.03T^{-1/2}_{IC8}L_{Edd}$) \citep{Begelman_1983}.   For Eddington Luminosity of $L_{Edd} = 4.28 \times 10^{45}~\mathrm{erg/s}$ and a bolometric luminosity ranging from $L_{bol} = 5.3 - 8 \times 10^{43}~\mathrm{erg/s}$ (see \S\ref{sec: Mdot_KE}), $L_{bol}/L_{cr} = 0.026-0.039$, far below the critical point of unity. The typical velocity from thermally driven outflows is within $\sim400~\mathrm{km/s}$ \citep{Waters_2021}. If warm absorbers originate far from the central engine in NGC 4151, thermal pressure may contribute to launching the most distant warm absorbers, as also suggested by the extended $\log\xi$ range and clumpy AMD structure discussed in \S\ref{sec:AMD} and \cite{Waters_2021}. On the other hand, UFO and VFO components reach velocities of $\sim 10^3 - 10^5~\mathrm{km~s^{-1}}$ with signs of continuing acceleration. These components may represent the early stages of clump formation in a smooth outflow that can enter the dynamical thermal instability zone and give rise to clumpy structures at large distances \citep{Waters_2022}. Therefore, it suggests that thermal pressure may contribute to launching or shaping slower, more extended components, like WAs, but is unlikely to be the dominant driving mechanism, especially for the highest-velocity phases of the wind.

Radiation force on lines is inefficient for highly ionized gas with $\log{\xi} > 3$, because the line force drops to zero \citep{Dannen_2019}.  All of the UFOs have ionization parameters that are too high to be driven by radiation pressure on lines.  Half of the WAs, and a few VFOs with low ionization parameters ($\log{\xi} < 3$) could be driven by radiation pressure on lines, at least in part.  More highly ionized components can still be driven by electron scattering pressure, where a direct proportionality between the outflow momentum rate, $\dot{p}_{out} \equiv \dot{M}_{out}v_{out}$, and the momentum flux of the radiation field, $\dot{p}_{rad} \sim L_{bol}/c$ is expected \citep{Gofford_2015}.  Our models make it possible to examine this possibility in detail:  the momentum flux of the radiation field for each detected outflow component is estimated using the corrected bolometric luminosity of $L_{bol} = 2L_{\mathrm{ion}}$. The ionization luminosity seen by each component ($L_{\mathrm{ion}}$) is calculated by multiplying the ``\texttt{lixi}'' output from ``\texttt{$pion$}'' by its corresponding value of $\xi$.  

\begin{figure}
\includegraphics[width=0.48\textwidth]{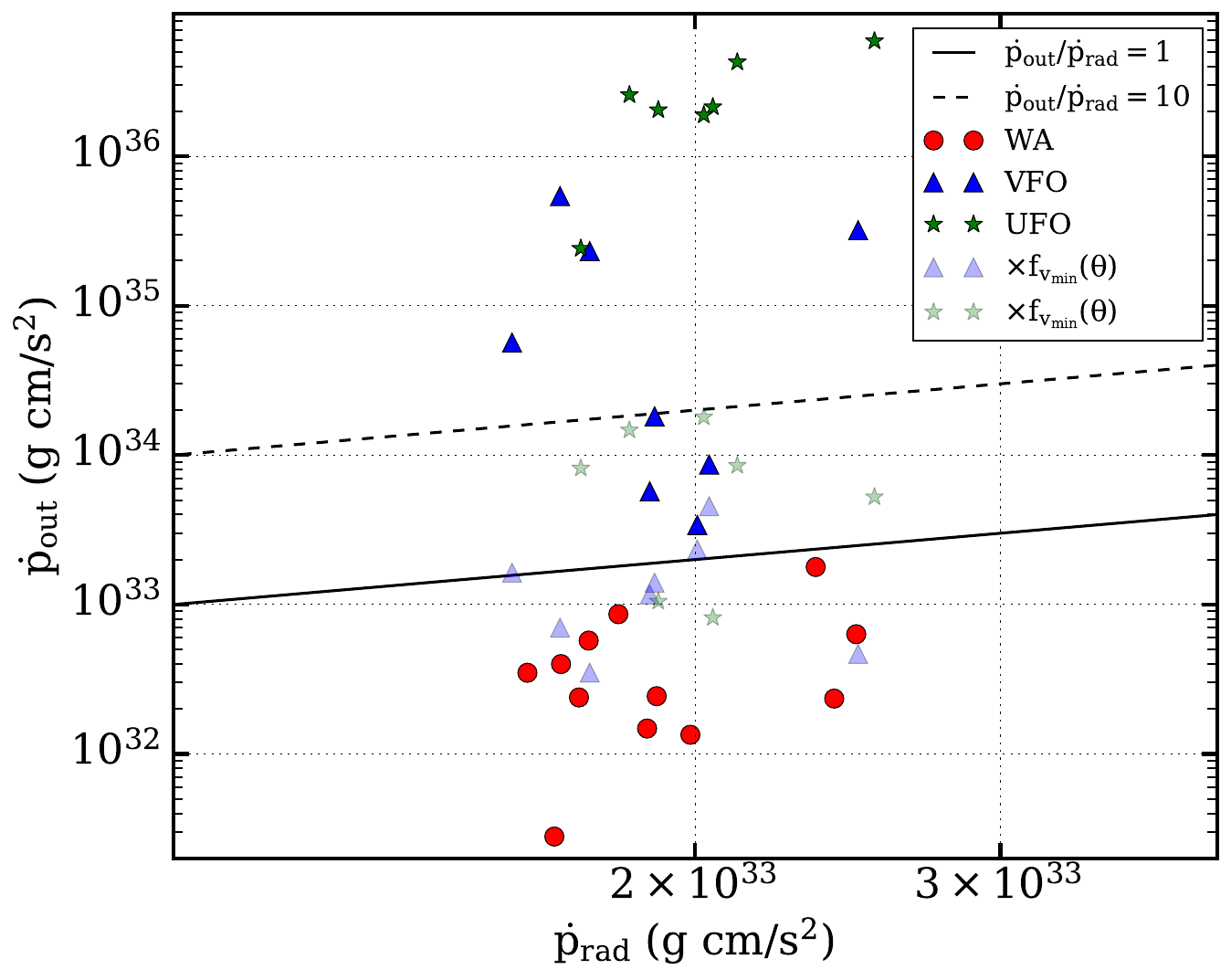}
\caption{Comparison of the outflow momentum rate ($\dot{p}_{out}$) the the momentum flux of the radiation field ($\dot{p}_{rad}$). The solid and dashed lines show the ratio of the two being 1 and 10, respectively. The transparent markers represent values multiplied by the lower limit of the volume filling factor.}
\label{fig: Pout}
\end{figure}

Figure \ref{fig: Pout} shows the momentum flux of the radiation field versus the outflow momentum rate for each wind component.  A simple OLS linear regression for the three types of outflows gives a slope of $2.56 \pm 2.03$, $-1.56 \pm 5.63$, and $6.41 \pm 2.28$ for the WAs, VFOs, and UFOs respectively (before correcting for filling factors), and slopes of $-0.04 \pm 2.44$ and $-0.84 \pm 4.38$ for VFOs, and UFOs after correcting by the minimum values of volume filling factor.  A simple linear relationship is excluded for the UFOs, but allowed for the VFOs and WAs, although the slopes are poorly constrained due to the narrow range of bolometric luminosity that is sampled.

The large uncertainties in the volume filling factor make the overall situation more complex. All of the UFOs and VFOs have outflow momentum beyond the momentum flux of the radiation field, beyond what electron scattering can transfer.  Even after considering the lowest volume filling factor, most UFOs still require additional driving mechanisms.  Only WAs and a few VFOs exhibit momentum ratios around unity (0.1 -- 10) that are consistent with radiative driving.  Further support for radiatively driven WAs comes from their velocity-ionization relationship in Figure \ref{fig: AMDvz}, where the best-fit relationship between the measured LOS velocity $v_z$ and ionization parameters $\xi$ are presented, using the same fitting method applied to the AMD analysis in Figure \ref{fig: AMD} discussed in \S\ref{sec:AMD}.  The results indicate that WAs exhibit the steepest slope $m = 0.48 \pm 0.19$, and $m = 0.24 \pm 0.19$ while excluding the low-significance cases, consistent with the proportion of $v \propto \xi^{1/2}$ that is expected from the linear proportionally of $\dot{p}_{out} \propto \dot{p}_{rad}$ for electron scattering pressure \citep{Tombesi_2013}.

In addition, the slope of $0.5$ in $\log{v}-\log{\xi}$ diagram is in agreement with momentum--conserving outflows where $\dot{p}_{out} = \dot{M}_{out}v \propto L_{\mathrm{ion}}v^2 / \xi$ is constant throughout the flow.  UFOs show a flatter slope, $m = 0.22 \pm 0.53$ ($m = 0.04 \pm 0.79$, when excluding cases with low statistical significance).  VFOs display an almost flat and even slightly negative slope of $m = -0.39 \pm 0.44$ ($m = - 0.07 \pm 0.33$). These flat slopes across individual wind components suggest that each type of outflow maintains a nearly constant LOS velocity over a range of ionization parameters, contrary to the predictions of radiatively driven winds.  This may hint a more complex geometry of the disk outflow that will be discussed in \S\ref{sec: structure}.

In summary, our analysis strongly favors a hybrid wind-launching scenario: UFOs and fast VFOs may be magnetocentrifugal flows, whereas slower WAs are likely to be driven radiatively.  Neither mechanism can act exclusively; instead, they may act jointly, resulting in a stratified wind structure, with magnetic fields dominating the inner, high-velocity regions, and radiation driving significant parts of the slower, outer layers.

\subsection{The geometry and structure of the disk wind} \label{sec: structure}
A spherically symmetric radial flow is expected to have a density profile of $n \propto r^{-2}$, whereas the profile derived from the AMD falls in the range $n \propto r^{-1.3}-r^{-1.7}$  for a broad array of considerations.  This fact, and the clear dominance of absorption lines over emission lines, signals that the outflow is not spherically symmetric.  Our results suggest a complex disk--wind geometry, spanning orders of magnitude in radius and ionization, and with variable projected velocities, outflow rates, and kinetic power.  In addition, the significant variations in the wind parameters across five observations suggest considerable asymmetry and clumpiness. The presence of blue-shifted emission features and the inference that not all winds escape to infinity add to the complexity of the picture that emerges.

The terminal velocity of wind launched from a disk is expected to follow Keplerian scaling, $v \propto r^{-1/2}$, though the launching physics may differ (e.g., \citealt{Blandford_Payne_1982, Castor_1975, Proga_1999}). In magnetocentrifugal winds, this scaling arises because gas is accelerated along magnetic field lines anchored on the disk \citep{Blandford_Payne_1982}.  In radiatively driven winds, the same scaling emerges from numerical calculations \citep{Proga_1999}.

\begin{figure*}
\includegraphics[width=\textwidth]{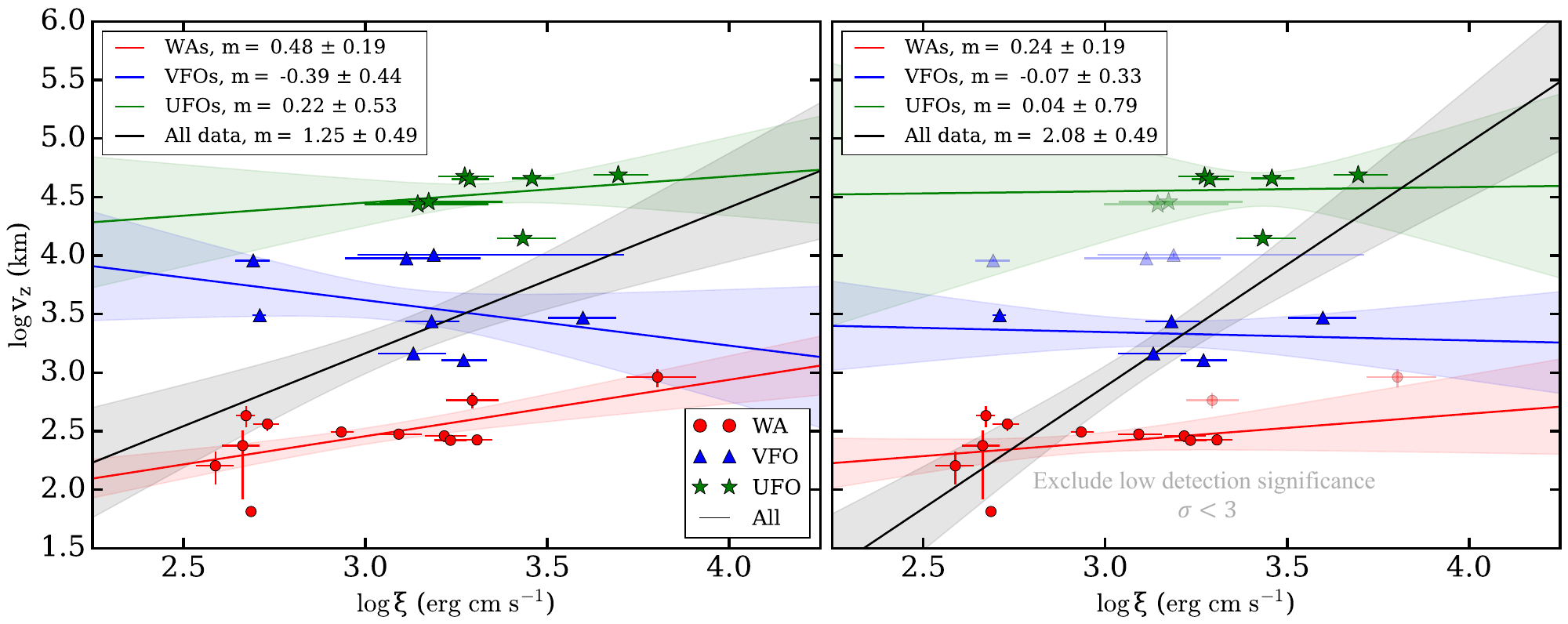}
\caption{The measured wind velocity along the line-of-sight ($v_z$) as a function of ionization parameters ($\xi$) for different wind types (WAs, VFOs, UFOs). The best-fit lines and their slopes $m$ for different types of outflows are shown, with shaded regions indicating the 1$\sigma$ confidence levels. The right panel excludes the outflows with detection significance below 3$\sigma$}
\label{fig: AMDvz}
\end{figure*}

In the context of the AMD, $n(r) \propto r^{\alpha}$ and $\xi \propto 1/nr^2$, so velocity is related to ionization via $v \propto \xi^{1/(4-2\alpha)}$. For a Blandford-Payne wind with $\alpha$ = 1.5, we expect a relation of $v \propto \xi$. In addition, from the measured density profiles ranging from $\alpha~=~1.4-1.7$, the expected slope of $\log{v} - \log{\xi}$ is $m~=~0.8-1.7$. However, the observed relation in Figure \ref{fig: AMDvz} deviates from this prediction, especially for the highly ionized and rapidly moving UFOs, and VFOs. Inferences based on the AMD rely on its underlying assumption that a single driving mechanism dominated at all radii and scale heights; this may be incorrect.  However, it is a simple assumption and therefore a good starting point, and it is worth considering whether a simple physical mechanism could flatten the velocity-ionization trend. 

\begin{figure*}
\includegraphics[width=\textwidth]{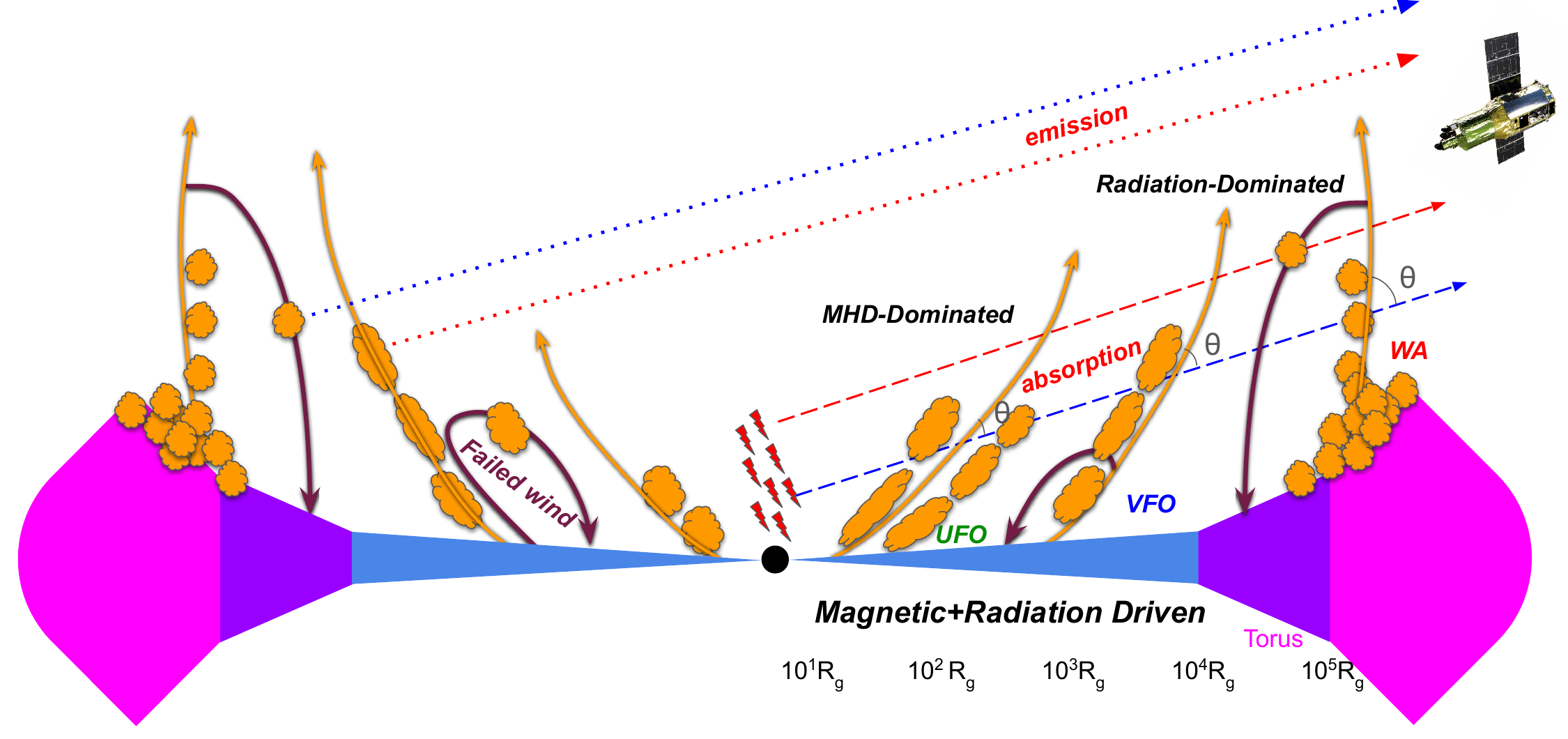}
\caption{Schematic illustration of proposed disk-wind geometry in NGC 4151. The wind exhibits multilayered, asymmetric, and clumpy structures. The continuous disk outflows exhibit a density profile $n \propto r^{1.5}$, consistent with Blandford-Payne MHD winds. Close to the central engine, Ultra Fast Outflows (UFOs) are primarily magnetohydrodynamically (MHD) driven due to their high ionization and velocity, while Warm Absorbers (WAs) at larger radii are predominantly driven by radiation pressure. The intermediate region containing Very Fast Outflows (VFOs) likely reflects a transition zone influenced by both radiation and magnetic mechanisms. The dashed lines represent the line-of-sight to the center engines, passing through the outflowing gas where the absorption features get created. The dotted lines represent the line-of-sight to the outflowing gas at the far side of the disk, where the emission features get created. The blue and red colors of the line-of-sight represent blue and red shifts of the absorption/emission features. As the winds propagate outwards, the angle ($\theta$) between the outflow and our line-of-sight is expected to increase, flattening the observed velocity-ionization slope. Failed winds (represented by downward arrows) indicate clumps of gas that are unable to achieve escape velocity and subsequently fall back onto the disk, producing observed blue-emission features.}
\label{fig: cartoon}
\end{figure*}

\begin{figure*}
\includegraphics[width=\textwidth]{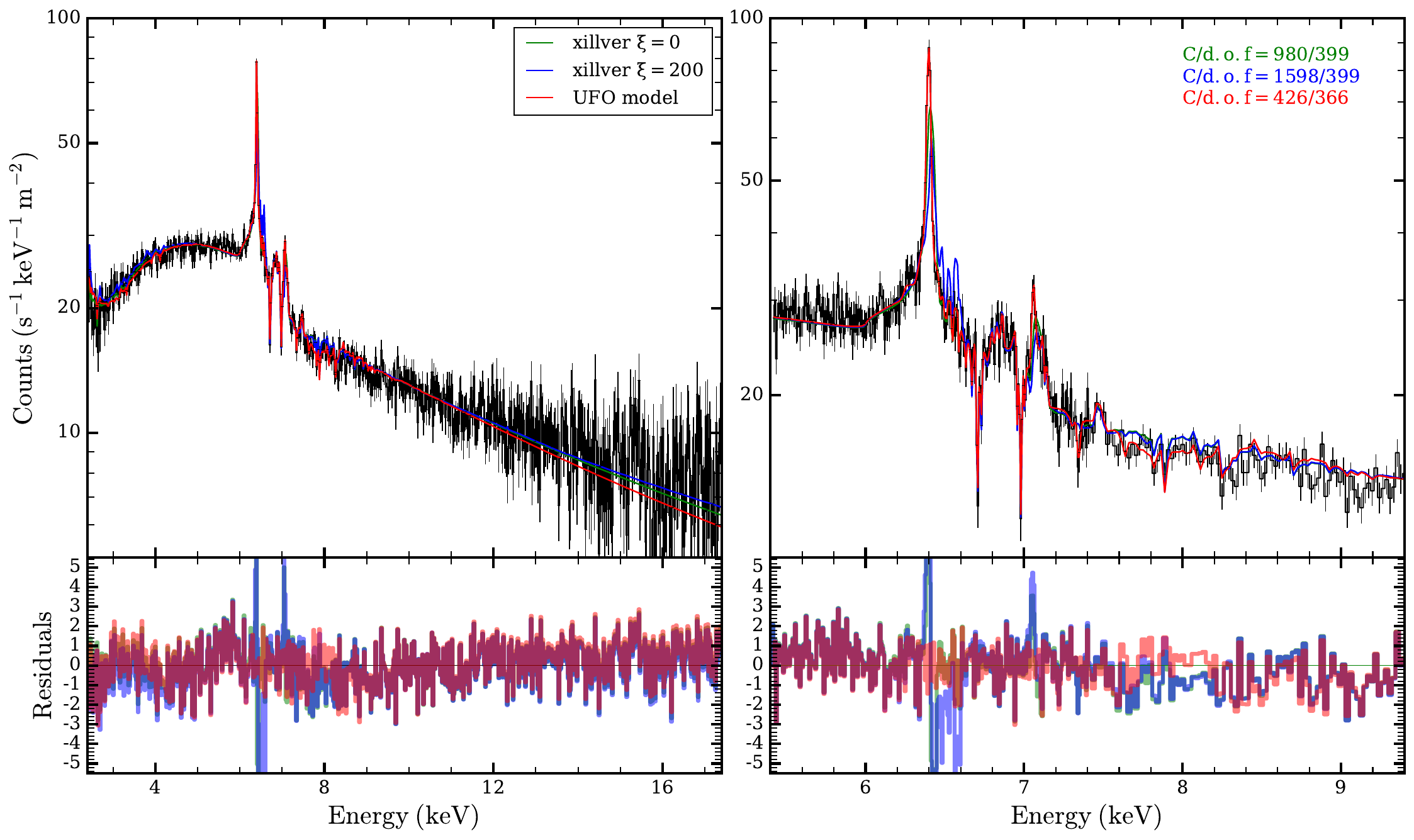}
\caption{XRISM observation of NGC 4151 on June 15 (Obs. 4) fitted with two different \texttt{xillver} models with $\xi = 0$ (green curve) and $\xi = 200$ (blue curve) with a broadband view of 2.4--17.4 keV (LEFT) and zoom in view of 5.4--10.4 keV (RIGHT). For comparison, the best-fit model with UFO components is shown as red curve. The lower panel shows the residuals ((data - Model) / Model) for different models. Both \texttt{xillver} models left strong residuals near 8 keV.}
\label{fig: june15xillver}
\end{figure*}

A possible explanation is a systematic change in the wind inclination angle relative to our line-of-sight, as illustrated schematically by the angle $\theta$ in Figure \ref{fig: cartoon}. The measured line-of-sight velocity is related to the wind velocity by $v_z = v \cos{\theta}$. In the self-similar models of a BP wind, the launching angle is initially fixed at the base along the magnetic field lines, but as the wind propagates outward, it can diverge, potentially leading to an increase in $\theta$ with radius. This collimated structure occurs as the field lines bend towards the rotation axis by magnetic tension \citep{Blandford_Payne_1982}.  If the wind gradually bends away from the equatorial plane, such that $\theta(r)$ follows a positive power-law relation with radius, then the observed line-of-sight velocity systematically decreases with radial distance from the black hole, progressively flattening the expected slope for each wind component. This interpretation may be independently supported by the need for larger angles to reconcile the large volume-filling factors of further WAs (\S\ref{sec: r_fv}).  These inferences are broadly consistent with a ``bowl--shaped'' geometry inferred from broadband reverberation mapping in NGC 4151 and other AGN \citep{Starkey_2023, Edelson_2017, Edelson_2024}.

A potential alternative is that these inferences are flawed because the observed velocities do not reflect the launching radius, at least not for all outflow categories.  Above, we have noted that WAs may be failed winds, launched from relatively small radii with velocities that fall below the local escape velocity.  Although the errors are considerable, the slopes of the $\log{v} - \log{\xi}$ relation nominally differs between wind categories.  Even some of the VFOs may be close to the threshold of escaping to infinity versus returning to the system.  Certainly, the blue-shifted emission components with velocities exceeding $v \geq 5000~{\rm km}~{\rm s}^{-1}$ in Obs. 4 and Obs. 5 may represent failed wind components.  

Such failed winds have critical implications for our understanding of disk--wind interactions, as they represent a mechanism for redistributing mass and energy back to the accretion disk.  Circulating flows may feed the wind-launching region, resulting a dynamic interplay outflow-inflow geometry \citep{Kurosawa_Proga_2009}. The rapid variability observed in these blue-shifted emission features may indicate that we may be observing gas streams launched at various angles above of below our line-of-sight, causing temporal changes in observed emission properties.  Variable axial asymmetry in the outflow (and, indeed, the full, coupled, inflow--outflow system) may be another important complicating factor.  

The broad picture that emerges from the five PV observations of NGC 4151 is of a complex and dynamic disk-wind structure, where different wind phases coexist, influenced by various physical mechanism and geometric effects, shaping the observed emission and absorption features. A schematic illustration summarizing this intricate geometry is shown in Figure \ref{fig: cartoon}.  Future observation may be able to track the evolution of gas as it moves radially outward, or be able to locate wind components through internal variability or responses to the central engine, and may even be able to trace asymmetries as they pass across our line of sight to the central engine.

\subsection{Limitations}
While this work provides a comprehensive analysis of the high-ionization outflows observed in NGC 4151 with XRISM, several limitations remain:

1. Degeneracy in Absorber Layering: The ordering of the absorption layers in the photoionization model is not uniquely constrained. For example, swapping component\#1 and component\#3 in Obs. 3 increases the $C$-value by only 0.22, which is not statistically significant. While the detection significance of individual layers remains robust, the geometric ordering and laying sequence of some components cannot be reliably inferred from spectral fitting alone.

2. Limited sensitivity to low-ionization Gas: Previous studies, such as \cite{Detmers_2011}, have revealed multiple discrete layers of WAs spanning a broader ionization range, particularly at lower values ($\log\xi = 1 - 3$), which lie outside the main sensitivity range of this XRISM analysis focused on Fe K-shell transitions. Additional analysis of lower-energy bands is needed to fully characterize the presence and structure of low-ionization outflows. 

3. Incomplete Sampling for Variability and Structure: Although we detect multi-layered outflows and estimated their launching radii and densities, the sparse cadence and limited number of observations restrict our ability to track temporal evolution and to confidently establish long-term variability or correlations between parameters (e.g., between $L_{\mathrm{ion}}$, $\xi$, $v_z$, and $N_H$). More frequent observations and time-resolved spectroscopy will be essential to refine the structure and dynamics of the wind, and will be analyzed in our future work.

4. While our analysis does not directly determine why some components are failed winds while others escape to infinity, their simultaneous presence points to a rich diversity in wind dynamics. It remains an open question whether these components are especially massive and/or dense, or if they have collided with other wind components or ambient gas. These possibilities and many others are plausible.

\section{Conclusion} \label{sec: summary}
We have analyzed five XRISM observations on Seyfert 1.5 AGN NGC 4151 and unraveled the rich complexities of multiple layers of wind outflows across the observations. The main results can be summarized below:

1. Multiple types of disk winds -- including warm absorbers (WAs), very fast outflows (VFOs), and ultra-fast outflows (UFOs) -- can coexist simultaneously in NGC 4151.  Up to six photoionized absorption components are identified in a single observation. The observed positive correlation of column density and  ionization parameters confirms that the most ionized gas likely carries the bulk of the mass in black hole outflows. The rich layering of wind phases highlights the complexity of the disk-wind structure and supports the existence of a multi-phase, stratified outflow originating across a wide range of radii. 

2. UFOs are likely detected in all observations. These include broad UFOs in 4/5 ($80\%$) of the observations with consistent velocities around $0.2c$,  moderate UFOs at $0.05c$ in one observation, and narrow, lower ionized UFOs at $0.1c$ in 2/5 ($40\%$) of the observations. UFOs can be persistent with velocities potentially exceeding that of the known jet in NGC 4151 \citep{Ulvestad_2005}.

3. The nature of WAs remains uncertain.  WAs might be observed far from their launching points as they are observed to be stable within a week but vary in a month.  But, even this time interval implies radii that are orders of magnitude smaller than equating their velocities with a local escape velocity.  WA components may be observed as they flow outward but ultimately represent failed winds.  

4. Transient, highly blue-shifted emission components are detected in $2/5$ or $40\%$ of the observations.   The significant variability observed on timescales as short as one week is comparable to the light crossing time to the Broad Line Region (BLR), and suggests that these features may be anisotropic, failed streams of gas seen on the far side of the disk. 

5. VFOs and UFOs present higher variability, observed over timescales as short as a week, suggests that they originate on scales comparable to or potentially much smaller than the BLR.

6. All the outflows possess mass outflow rates comparable to or greater than the mass accretion rate, suggesting that even slow winds and/or those launched from larger radii can have a real effect on how much gas reaches the event horizon.  

7. All UFOs and most VFOs have a kinetic luminosity that exceeds $0.5\%$ of the Eddington luminosity -- the threshold for AGN feedback.  Even when assuming the lowest plausible volume fitting factors, two UFOs still exceed this threshold, marking them as strong candidates to halt star formation in the galactic bulge. 

8. All wind components above a basic $> 3\sigma$ statistical threshold give rise to an AMD and density profile that is consistent with magnetocentrifugal driving.  Contributions from radiative driving are also evident, particularly in lower-ionization WAs.

\begin{acknowledgments}
We thank Jelle de Plaa, Liyi Gu, and Chen Li for their helpful conversation about SPEX. We are grateful to the anonymous referee for a timely review and pointing us to relevant theoretical work on the formation of clumpiness and stratification by thermal instability.

\end{acknowledgments}

\bibliography{main}{}
\bibliographystyle{aasjournal}

\clearpage

\begin{deluxetable*}{cclllll} 
\tabletypesize{\footnotesize}
\label{table:parameters}
\tablecaption{Best-fit Parameters and Detection Significance (D.S) for the outflow components.}
\tablewidth{0pt}
\tablehead{
\colhead{Components} & \colhead{Parameters} & \colhead{Obs. 1} & \colhead{Obs. 2} & \colhead{Obs. 3} & \colhead{Obs. 4} & \colhead{Obs. 5}
}
\startdata
    $pion_{\#1}$ & $N_H$ ($\times 10^{24}~\rm cm^{-2}$) & $0.015^{+0.003}_{-0.002}$ & $0.026^{+0.003}_{-0.003}$ & $0.014^{+0.002}_{-0.002}$ & $0.013^{+0.002}_{-0.002}$ & $0.012^{+0.002}_{-0.002}$\\
                       &$\log \xi$ ($\mathrm{erg~cm~s^{-1}}$) & $2.663^{+0.047}_{-0.057}$ & $2.672^{+0.024}_{-0.028}$ & $2.710^{+0.017}_{-0.020}$ & $2.588^{+0.050}_{-0.054}$ & $2.731^{+0.032}_{-0.040}$\\
                        & $\sigma_v$ ($\mathrm{km~s^{-1}}$) & $270^{+171}_{-123}$ & $395^{+74}_{-66}$ & $179^{+58}_{-50}$ & $100^{+85}_{-0}$ & $100^{+44}_{-0}$\\
                        & $v_z$ ($\mathrm{km~s^{-1}}$) & $-238^{+155}_{-83}$ & $-429^{+90}_{-85}$ & $^*-3092^{+46}_{-47}$ & $-160^{+49}_{-51}$ & $-364^{+49}_{-43}$\\
    D.S ($\Delta$AIC) & & 3.6$\sigma$ (-14) & 4.8$\sigma$ (-25) & 3.0$\sigma$ (-6.5) & 4.3$\sigma$ (-17) & 4.0$\sigma$ (-16)\\
    \hline
    $pion_{\#2}$ & $N_H$ ($\times 10^{24}~\rm cm^{-2}$) & $0.059^{+0.009}_{-0.008}$ & $0.062^{+0.007}_{-0.007}$ & $0.041^{+0.005}_{-0.004}$ & $0.109^{+0.008}_{-0.008}$ & $0.085^{+0.007}_{-0.007}$\\
                       & $\log \xi$ ($\mathrm{erg~cm~s^{-1}}$) & $3.092^{+0.063}_{-0.056}$ & $3.217^{+0.060}_{-0.054}$ & $2.934^{+0.034}_{-0.029}$ & $3.307^{+0.043}_{-0.042}$ & $3.234^{+0.044}_{-0.044}$\\
                        & $\sigma_v$ ($\mathrm{km~s^{-1}}$) & $100^{+14}_{-0}$ & $183^{+31}_{-26}$ & $129^{+19}_{-19}$ & $142^{+17}_{-16}$ & $155^{+18}_{-17}$\\
                        & $v_z$ ($\mathrm{km~s^{-1}}$) & $-298^{+28}_{-29}$ & $-288^{+32}_{-30}$ & $-311^{+24}_{-23}$ & $-266^{+22}_{-21}$ & $-264^{+23}_{-23}$\\
    D.S ($\Delta$AIC) & & 3.5$\sigma$ (-13) & 6.3$\sigma$ (-43) & 7.8$\sigma$ (-57) & 6.9$\sigma$ (-43) & 6.1$\sigma$ (-37)\\
    \hline
    $pion_{\#3}$ & $N_H$ ($\times 10^{24}~\rm cm^{-2}$) & $0.032^{+0.007}_{-0.006}$ & $0.075^{+0.008}_{-0.008}$ & $0.030^{+0.002}_{-0.002}$ & $0.020^{+0.005}_{-0.004}$ & $0.010^{+0.003}_{-0.002}$\\
                       & $\log \xi$ ($\mathrm{erg~cm~s^{-1}}$) & $3.294^{+0.073}_{-0.072}$ & $3.803^{+0.106}_{-0.086}$ & $2.686^{+0.012}_{-0.013}$ & $3.270^{+0.065}_{-0.062}$ & $3.132^{+0.090}_{-0.097}$\\
                        & $\sigma_v$ ($\mathrm{km~s^{-1}}$) & $274^{+90}_{-58}$ & $314^{+167}_{-108}$ & $406^{+52}_{-46}$ & $223^{+58}_{-51}$ & $139^{+74}_{-39}$\\
                        & $v_z$ ($\mathrm{km~s^{-1}}$) & $-579^{+88}_{-93}$ & $-915^{+169}_{-158}$ & $-65^{+69}_{-65}$ & $^*-1277^{+63}_{-59}$ & $^*-1457^{+55}_{-56}$\\
    D.S ($\Delta$AIC) & & 2.1$\sigma$ (-2.7) & 1.5$\sigma$ (0.5) & 4.5$\sigma$ (-19) & 3.7$\sigma$ (-12) & 4.1$\sigma$ (-16)\\
    \hline
    $pion_{\#4}$  & $N_H$ ($\times 10^{24}~\rm cm^{-2}$) & $0.003^{+0.003}_{-0.002}$ & $0.004^{+0.009}_{-0.002}$ & $0.006^{+0.002}_{-0.002}$ & $0.039^{+0.008}_{-0.008}$ & $0.012^{+0.003}_{-0.003}$\\
                        & $\log \xi$ ($\mathrm{erg~cm~s^{-1}}$) & $3.113^{+0.204}_{-0.170}$ & $3.188^{+0.523}_{-0.209}$ & $2.692^{+0.045}_{-0.050}$ & $3.598^{+0.091}_{-0.096}$ & $3.182^{+0.077}_{-0.072}$\\
                        & $\sigma_v$ ($\mathrm{km~s^{-1}}$) & $100^{+125}_{-0}$ & $411^{+455}_{-200}$ & $182^{+180}_{-82}$ & $400^{+189}_{-138}$ & $231^{+106}_{-79}$\\
                        & $v_z$ ($\mathrm{km~s^{-1}}$) & $-9445^{+116}_{-133}$ & $-10123^{+465}_{-404}$ & $-9071^{+106}_{-146}$ & $-2946^{+173}_{-186}$ & $-2742^{+93}_{-74}$\\
    D.S ($\Delta$AIC) & & 0.6$\sigma$ (4.8) & 0.9$\sigma$ (3.4) & 2.1$\sigma$ (-2.1) & 4.1$\sigma$ (-15) & 3.1$\sigma$ (-9)\\
    \hline
    $pion_{\#5}$    & $N_H$ ($\times 10^{24}~\rm cm^{-2}$) & ... & ... & $0.070^{+0.008}_{-0.008}$ & $0.007^{+0.010}_{-0.003}$ & $0.009^{+0.004}_{-0.003}$\\
                        & $\log \xi$ ($\mathrm{erg~cm~s^{-1}}$) & ... & ... & $3.433^{+0.090}_{-0.072}$ & $3.144^{+0.194}_{-0.147}$ & $3.174^{+0.204}_{-0.136}$\\
                        & $\sigma_v$ ($\mathrm{km~s^{-1}}$) & ... & ... & $2203^{+486}_{-468}$ & $306^{+1823}_{-180}$ & $583^{+374}_{-423}$\\
                        & $v_z$ ($\mathrm{km~s^{-1}}$) & ... & ... & $-14022^{+636}_{-736}$ & $-27402^{+331}_{-1067}$ & $-28753^{+288}_{-331}$\\
    D.S ($\Delta$AIC) & & ... & ... & 3.9$\sigma$ (-14) & 2.1$\sigma$ (-1.7) & 2.3$\sigma$ (-3.1)\\
    \hline
    $pion_{\#6}$    & $N_H$ ($\times 10^{24}~\rm cm^{-2}$) & $0.126^{+0.010}_{-0.010}$ & $0.045^{+0.007}_{-0.006}$ & ... & $0.073^{+0.007}_{-0.007}$ & $0.154^{+0.009}_{-0.008}$ \\
                        & $\log \xi$ ($\mathrm{erg~cm~s^{-1}}$) & $3.458^{+0.062}_{-0.056}$ & $3.273^{+0.081}_{-0.073}$ & ... & $3.287^{+0.053}_{-0.050}$ & $3.695^{+0.082}_{-0.067}$\\
                        & $\sigma_v$ ($\mathrm{km~s^{-1}}$) & $5000^{+0}_{-2198}$ & $5000^{+0}_{-910}$ & ... & $5000^{+0}_{-211}$ & $5000^{+0}_{-388}$\\
                        & $v_z$ ($\mathrm{km~s^{-1}}$) & $-45633^{+1352}_{-1311}$ & $-47444^{+2142}_{-1783}$ & ... & $-44924^{+2527}_{-1424}$ & $-48836^{+2197}_{-2270}$\\
    D.S ($\Delta$AIC) & & 5.3$\sigma$ (-29) & 4.0$\sigma$ (-17) & ... & 6.0$\sigma$ (-33) & 7.6$\sigma$ (-56)\\
    \hline
    $pion^{emis\#2}$& $v_z$ ($\mathrm{km~s^{-1}}$) & $500^{+0}_{-58}$ & ... & $382^{+55}_{-75}$ & $692^{+200}_{-220}$ & $921^{+79}_{-149}$\\
                        & $\Omega$ & $0.08^{+0.02}_{-0.03}$ & ... & $0.13^{+0.03}_{-0.02}$ & $0.09^{+0.02}_{-0.02}$ & $0.11^{+0.02}_{-0.02}$\\
    $vgau$         & $\sigma$ ($\mathrm{km~s^{-1}}$) & $246^{+45}_{-48}$ & ... & $71^{+137}_{-71}$ & $426^{+27}_{-27}$ & $315^{+39}_{-39}$\\
    D.S ($\Delta$AIC) & & 1.1$\sigma$ (2.0) & ... & 2.9$\sigma$ (-6.5) & 2.2$\sigma$ (-2.7) & 3.1$\sigma$ (-9)\\
    \hline
    $pion^{blue-emis}$ & $N_H$ ($\times 10^{24}~\rm cm^{-2}$) & ... & ... & ... & $0.1^-$ & $0.1^-$\\
                        & $\log \xi$ ($\mathrm{erg~cm~s^{-1}}$) & ... & ... & ... & $3.155^{+0.064}_{-0.061}$ & $2.799^{+0.016}_{-0.020}$\\
                        & $\sigma_v$ ($\mathrm{km~s^{-1}}$) & ... & ... & ... & $100^-$ & $100^-$\\
                        & $v_z$ ($\mathrm{km~s^{-1}}$) & ... & ... & ... & $-7149^{+91}_{-95}$ & $-5494^{+240}_{-236}$\\
                        & $\Omega$ & ... & ...  & ... & $0.106^{+0.024}_{-0.021}$ & $0.165^{+0.015}_{-0.015}$\\
    $vgau$         & $\sigma$ ($\mathrm{km~s^{-1}}$) & ... & ...  & ... & $269^{+95}_{-72}$ & $681^{+178}_{-117}$\\
    D.S ($\Delta$AIC) & & ... & ... & ... & 5.3$\sigma$ (-26) & 5.5$\sigma$ (-30)\\
    \hline
    $pow$            & $\Gamma$ & $1.64^{+0.005}_{-0.016}$ & $1.72^{+0.002}_{-0.026}$ & $1.57^{+0.036}_{-0.028}$ & $1.70^{+0.019}_{-0.005}$ & $1.69^{+0.033}_{-0.029}$ \\
                        & $\mathrm{Norm}~\times 10^7$ & $1.77^{+0.024}_{-0.052}$ & $2.96^{+0.018}_{-0.135}$ & $1.52^{+0.144}_{-0.063}$ & $2.27^{+0.075}_{-0.034}$ & $2.10^{+0.178}_{-0.009}$\\ 
    \hline
    $hot$           & $N_H$ ($\times 10^{24}~\rm cm^{-2}$) & $0.19^{+0.01}_{-0.01}$ & $0.15^{+0.03}_{-0.02}$ & $0.14^{+0.02}_{-0.02}$ & $0.21^{+0.01}_{-0.01}$ & $0.15^{+0.02}_{-0.02}$\\
    & $f_{cov}$ & $0.86^{+0.01}_{-0.01}$ & $0.85^{+0.04}_{-0.04}$ & $0.87^{+0.01}_{-0.03}$ & $0.89^{+0.03}_{-0.03}$ & $0.86^{+0.03}_{-0.03}$\\
    \hline
    $C$/d.o.f             & & 380/374 & 371/374 & 413/376 & 426/366 & 404/367\\
    Expected $C$-values     & & $399\pm28$ & $399\pm28$ & $400\pm28$ & $399\pm28$ & $401\pm28$ \\
\enddata
\tablecomments{The total model written in SPEX language, in the order from source to us for the $pion$ absorbers:
\texttt{$((bb + (pow + 3 \times (mytorus * spei)) * etau_{low} * etau_{hi}) * line_{1} * line_{2} *  pion_{\#6} * pion_{\#5} * pion_{\#4} * pion_{\#3} * pion_{\#2} * pion_{\#1} + vgau*pion^{emis\#2} + vgau*pion^{blue-emis}) * hot * reds$} \\
$^*$ The VFO component in \#1--3 while most other observations has WAs in this layer. \\
}
\end{deluxetable*}

\end{document}